\documentclass[english,aps,prl,amsmath,amssymb,twocolumn,letterpaper,showpacs,superscriptaddress]{revtex4-1}

\usepackage{mathtools} 

\usepackage[normalem]{ulem} 

\usepackage{comment}
\usepackage{amsmath}
\usepackage{physics}
\usepackage{amssymb}
\usepackage{graphicx,bm}
\usepackage{makecell,multirow}
\usepackage{babel}
\usepackage{amstext}
\usepackage{esint}
\usepackage{cases}
\usepackage{makecell}
\usepackage{lineno}
\usepackage{soul}
\usepackage[caption=false]{subfig}
\usepackage{verbatim}
\usepackage[unicode=true,pdfusetitle,
bookmarks=true,bookmarksnumbered=false,bookmarksopen=false, breaklinks=false,pdfborder={0 0 1},backref=false,colorlinks=false]
 {hyperref}
 \hypersetup{
    colorlinks=true,
    linkcolor=red,
    citecolor=blue,
    filecolor=magenta,      
    urlcolor=blue
    }
\usepackage{pdfpages}	
\usepackage{pgffor}		
\usepackage{upgreek}
\usepackage{braket}

\makeatletter
\AtBeginDocument{\let\LS@rot\@undefined} 
\makeatother							 

\newcommand{\prlsection}[1]{ \textit{#1.---}}
\begin{document}

\title{Quantum statistics and self-interference in extended colliders 
}

\author{Sai Satyam Samal}
\affiliation{Department of Physics and Astronomy, Purdue University, West Lafayette, Indiana 47907, USA}

\author{Smitha Vishveshwara}
\affiliation{Department of Physics, University of Illinois Urbana-Champaign, Urbana, Illinois 61801, USA}

\author{Yuval Gefen}
\affiliation{Department of Condensed Matter Physics, The Weizmann Institute of Science, Rehovot 76100, Israel}

\author{Jukka I. V\"{a}yrynen}
\affiliation{Department of Physics and Astronomy, Purdue University, West Lafayette, Indiana 47907, USA}

\date{\today}
\begin{abstract}
Collision of quantum particles remains an effective way of probing their mutual statistics. Colliders based on quantum point contacts in quantum Hall edge states have been successfully used to probe the statistics of the underlying quantum particles. Notwithstanding the extensive theoretical work focusing on point-like colliders, when it comes to experiment, the colliders are rarely point-like objects and can support a resonant level or multiple tunneling points. We present a study of a paradigmatic extended (non-point-like) fermionic collider (and an extension to bosonic colliders).  As with particle interferometers, in an extended collider there is an infinite number of trajectories for any single or multi-particle event. Self-interference of the former can lead to an apparent bunching of fermions when we compare the cross-current correlator with a classical benchmark representing two colliding beams. In view of this apparent bunching behavior of fermions, we identify an experimentally accessible current correlator which reveals the true mutual statistics of fermions.  
\end{abstract}

\maketitle

     \prlsection{Introduction} Braiding statistics constitutes a fundamental
     property of quantum particles and encodes how the wave function changes under the adiabatic encircling of one particle by another~\cite{doi:10.1142/5752}. While fermionic and bosonic statistics primarily underlie observable quantum many-body physics, two-dimensional platforms, e.g. 2D electron gas (2DEG) setups, may accommodate anyons exhibiting intermediate fractional statistics~\cite{leinaas_theory_1977,PhysRevLett.49.957,doi:10.1142/5752,rao2016introduction,RevModPhys.80.1083}. Fractional quantum Hall phases provide canonical platforms wherein such exotic quasi-particles can be manipulated~\cite{PhysRevLett.48.1559,girvin1999quantum,PhysRevLett.53.722}. Interference-based theoretical proposals for pinpointing quantum statistics~\cite{PhysRevB.55.2331,PhysRevB.74.045319,PhysRevB.76.085333,BONDERSON20082709,PhysRevB.105.165310,PhysRevB.108.L241302} have led to a surge of experimental realizations~\cite{Nakamura_2020,nakamura_aharonovbohm_2019,nakamura_impact_2022,PhysRevX.13.041012,Kundu_2023,neder_interference_2007,ronen_aharonovbohm_2021,kim_aharonovbohm_2024,werkmeister_strongly_2024,werkmeister2024anyonbraidingtelegraphnoise,PhysRevLett.108.256804,samuelson2024anyonicstatisticsslowquasiparticle}. Complementary and arguably simpler methods avoiding the use of interference in probing quantum statistics, also being actively explored, are based on so-called ``colliders'', where incoming quantum particles experience statistical interaction and are then detected by measuring electrical currents. Cross-current correlations~\cite{Blanter_2000} in Hanbury-Brown and Twiss settings are therein adapted to quantum Hall edge states~\cite{PhysRevLett.86.4628,GUYON2002697,PhysRevLett.91.196803,PhysRevLett.95.176402,PhysRevB.74.155324,PhysRevLett.109.106802,PhysRevB.88.235415,PhysRevLett.116.156802}. Avoiding the idiosyncratic features of anyonic phenomenology, we will show that even in the seemingly established cases of fermions and bosons, the subtleties of collider dynamics may mask the quantum statistics of the particles involved.

     With point-like colliders, the statistics of quantum particles is determined by measuring current auto-correlations or cross-correlations~\cite{Blanter_2000,Lee_2022,Han_2016,doi:10.1126/science.aaz5601,PhysRevX.13.011031,Lee_2023,bocquillon_coherence_2013,PhysRevLett.116.156802,PhysRevB.46.12485}. To determine the emergence of excess or deficiency of particle bunching, one needs to compare these correlations to a benchmark~\cite{Lee_2022,Han_2016,PhysRevLett.116.156802,PhysRevB.88.235415,PhysRevB.99.045430,PhysRevLett.109.106802}. There are two candidate benchmarks which produce identical results in the case of a point-like collider. One is the corresponding cross-current correlator resulting from the collision of non-interacting, distinguishable classical particles~\cite{Blanter_2000,PhysRevB.99.045430}. The second is the reducible  part of the cross-current correlator~\footnote{Here reducible refers to all the single particle self-interference contributions to the cross/auto-correlations functions.} which incorporates all the single-particle contributions to the cross-current correlations  (including interference effects). Interestingly, the latter benchmark  is equivalent to  the cross-current correlator associated with classical waves. Following this guiding principle,  Laughlin anyons  have been shown to give rise to boson-like bunching, quantitatively  intermediate between bosons and classical particles~\cite{PhysRevLett.91.196803,PhysRevLett.109.106802,PhysRevB.88.235415,PhysRevLett.116.156802,doi:10.1126/science.aaz5601,PhysRevX.13.011031,Lee_2023,bocquillon_coherence_2013,PhysRevLett.86.4628,GUYON2002697,PhysRevLett.95.176402,PhysRevB.74.155324,PhysRevLett.109.106802}.

Generalizing to non-point-like colliders,  we show here that the above two benchmarks lead to different results and that the choice of a correct benchmark is therefore crucial. While the interpretation of recent  experiments~\cite{doi:10.1126/science.aaz5601,PhysRevX.13.011031,Lee_2023,bocquillon_coherence_2013} draws on advances in the theory of point-like colliders~\cite{Lee_2022,Han_2016,PhysRevLett.116.156802}, in realistic experimental platforms, the colliders are almost invariably extended. They may host resonant levels, caused for instance by multiple tunneling bridges~\cite{PhysRevB.80.035319} or from interaction-induced quantum dot formation~\cite{PhysRevLett.134.076302}, observed in the quantum point contact (QPC) transmission curves~\cite{PhysRevX.13.011031,garg2025enhancedshotnoisegraphene}.  In such an ``extended collider”, multiple trajectories can connect the source and the drain, and interference effects may distort the statics-based results.

We find that employing the first benchmark to determine the boson-like or fermion-like nature of the particle-statistics is fundamentally flawed for extended colliders. One may  obtain 
an \emph{apparent} statistical transmutation~\cite{PhysRevB.99.045430}, i.e. fermions that seemingly tend to bunch. The origin of this observation relates to each particle’s self-interference contributions, having nothing to do with quantum statistics. Employing the second benchmark, i.e.,  subtracting
the reducible part from the cross-current correlator, one removes single-particle self-interference contributions, leading to a genuine quantum statistics witness. Thus, the second benchmark correctly captures the statistics of the colliding particles. In a point-like collider, single-particle self-interference terms are absent, which is why the two benchmarks lead to identical results. From a more practical point of view,  we note that the two benchmarks are accessible experimentally. By comparing them, we can readily conclude whether the collider at hand is indeed point-like (employing the two benchmarks yields the same result) or extended (the first bench mark indicates more bunching than the second one).

\begin{figure}[tbp!]
\centering
\includegraphics[width=1.0\columnwidth]{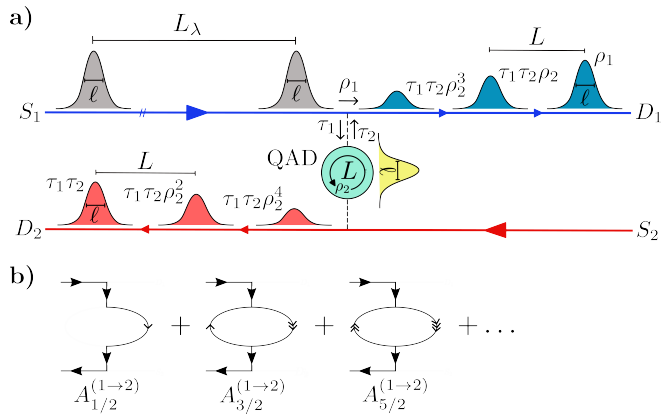}
\caption{
 \textbf{(a)}~Extended collider comprising two sources ($S_{1}$, $S_{2}$) and two detectors ($D_{1}$, $D_{2}$). A dilute beam having large separation $L_{\lambda}$ between wave packets of width $\ell$, emitted from $S_{1}$, can give rise to a quasi-bound state at the collision area, a chiral quantum anti-dot (QAD) of circumference  $L$. An incoming beam can tunnel into the QAD. As a result the outgoing beam comprises infinitely many wavelets of ever decreasing amplitude [determined by the junction tunneling amplitudes, Eq.~\eqref{eq:smatrix}], spatially separated by $L$. \textbf{(b)}~Processes that contribute to the amplitude to tunnel from $S_{1}$ to $D_{2}$. The arrows in each segment of the QAD represent the number of times this segment was visited before being emitted from the QAD in the direction of $D_{2}$. The processes involve half-integer number of windings, depicted by the subscript in the amplitude $A^{(1\to2)}_{n+\frac{1}{2}}$. Summing up these partial amplitudes facilitates evaluation of the ``anti-bunching probability" $P(11)_{\text{F/B}}$ for fermions/bosons, Sec.~III of SM, Ref.~\cite{Note2}.}
  \label{fig:extended_collider}
  \end{figure}

Our analysis here focuses on the fermionic collider while providing results for bosonic colliders where applicable. Such a focus offers two advantages: First, antibunching (bunching)
of fermions (bosons) is more extreme and ubiquitous than that of intermediate statistics of anyons. Second, we can avoid the complication due to  topological vacuum-bubbles braiding~\cite{Han_2016}, also known as time-domain braiding~\cite{PhysRevLett.116.156802,Lee_2022,PhysRevLett.131.186601}, thus facilitating a clear message concerning false  bunching of fermions.

Below, we first introduce a simple model of an extended collider that captures the relevant physics. We then consider a two-particle collision, demonstrating that the incorrect benchmark leads to apparent fermion bunching, leading us to identify the role of self-interference and define the proper statistical benchmark. Finally, we consider the case of many-particle colliding beams, and demonstrate that the results of our analysis can be tested in experiments by measuring dc (zero frequency) auto- and cross-current correlators.

\prlsection{Model and qualitative description of apparent bunching} Our model consists of two one-dimensional chiral channels emanating from sources $S_{1}$ and $S_{2}$, and terminating at the respective drains $D_{1}$ and $D_{2}$, where particles can be detected (e.g., by measuring currents). These channels are tunnel-coupled to the extended collider. The latter is modelled by a quantum anti-dot (QAD) encircled by a single chiral channel, cf. Fig.~\ref{fig:extended_collider}a. Arriving at the collider, particles (wave packets)  from the source ($S_{1}$ or $S_{2}$) can tunnel to and wind around the QAD an arbitrary number of times, before escaping to a detector drain ($D_{1}$ or $D_{2}$). For simplicity, we work with time-reversal symmetric and identical (see Sec.~II B of SM~\footnote{see Supplemental Material for details of the results discussed in the main text. Section I and II sets up the scattering formalism and discusses the differences between a point-like collider and an extended collider. Section III presents a toy model explaining false statistical transmutation signal. Section IV, V, VI calculates wave packet properties, presents a correct benchmark and discusses the dilute stochastic beam of emitted wave packets. Section VII and VII establishes the relation between the probabilities and current-current correlations and discusses $g^{2}-$function.} for non-identical) $S-$matrix at each channel-QAD junction (represented by a dashed line in  Fig.~\ref{fig:extended_collider}a), given by ($r>0$), 
\begin{align}\label{eq:smatrix}
        S= \left(\begin{array}{cc}
          \rho_{1} & \tau_{1}\\
         \tau_{2} & \rho_{2}
    \end{array}\right) =\left(\begin{array}{cc}
          re^{i\theta} & i\sqrt{1-r^{2}}\\
         i\sqrt{1-r^{2}} & re^{-i\theta}
    \end{array}\right) .
\end{align}

The origin of apparent fermion bunching in the extended collider setup can be simply understood by considering the limit of a  small edge-collider tunneling amplitude, $t=\sqrt{1-r^{2}}\ll 1$, and calculating the probability of a two-particle ``anti-bunching event", $P(11)_{\text{F}}$: the two colliding fermions end up each at a different drain. Here, apparent bunching-like behaviour corresponds to the inequality, $\mathcal{Q}_{\text{Cl;F}} = P(11)_{\text{F}} - P(11)_{\text{Cl}} \leq 0$, where comparison is with the corresponding benchmark  $\mathcal{B}_{1} = P(11)_{\text{Cl}}$ for classical particles. The latter probability, associated with two independent scattering events of classical particles, is equal to the sum of the probabilities of each of the contributing processes with no interference between different trajectories (or histories). 
For example, the event (to be denoted  $1 \to 2$) where a particle goes from $S_{1}$ to $D_{2}$, involves a particle transmitted through QAD, see Fig.~\ref{fig:extended_collider}b. The probability for this process is the sum of the squares of each amplitude in Fig.~\ref{fig:extended_collider}b, forming a geometric series $\mathcal{P}{(1\rightarrow 2)}_\text{Cl}=t^4 (1 + r^4 + \dots)=\frac{t^{2}}{2-t^{2}} \approx \frac{t^{2}}{2}$, for $t\ll 1$. Using conservation of probability, we get, $\mathcal{P}({1\rightarrow 1})_\text{Cl}=1-\frac{t^{2}}{2}$. Hence, for two uncorrelated and non-interacting classical particles one then obtains $P(11)_{\text{Cl}}=\mathcal{P}({1\rightarrow 1})_\text{Cl}^{2}+\mathcal{P}({1\rightarrow 2})_\text{Cl}^{2}=1-t^{2}$~\cite{Note2}.

Next, to obtain $P(11)_{\text{F}}$, we consider the leading quantum interference correction to the above result.  In the leading approximation, we restrict ourselves to consider interference only between trajectories whose winding number differ at most by 1. Such partial interference is relevant for the case where the spatial width $\ell$ of the wave packets is smaller than the size $L$ of the QAD, Fig.~\ref{fig:extended_collider}a. The ratio $\frac{\ell}{L}$ determines the degree of interference in the system. The limit $\frac{\ell}{L} \to 0$ produces the results for the classical particles (separate, non-overlapping wave packets do not interfere)~\footnote{By 'classical', we mean in this paper the complete lack of interference, still allowing for tunneling processes.}. In the opposite limit $\frac{\ell}{L}\gg1$, all trajectories interfere with each other.

For quantum particles, the correction due to self-interference of the single particle trajectories, Fig.~\ref{fig:extended_collider}b, modifies the single-particle tunneling probability to $\mathcal{P}{(1\rightarrow 2)}_\text{F} = \frac{t^{2}}{2} (1 + 2 c_1 \cos 2\alpha)$, where  $c_{1}\ll 1$ is a positive interference suppression factor ($c_1 = 0$ in the fully classical case) and $2 \alpha = k L - 2 \theta$ is the accumulated orbital phase. With the above result, we find $P(11)_{\text{F}} \approx \mathcal{P}{(1\rightarrow 1)_{\text{F}}^2} = 1-t^{2}(1+2c_{1} \cos 2\alpha)$, showing apparent bunching of fermions, $\mathcal{Q}_{\text{Cl;F}}=P(11)_{\text{F}}-P(11)_{\text{Cl}}=-2t^{2}c_{1}\cos 2\alpha \leq 0$  whenever $\cos 2\alpha\geq 0$. The bunching is largest near resonant tunneling~\cite{gordanQM}, where the interference enhancement is maximized, $2 \alpha/\pi = 0 \mod 2$. We note that this $t^2$-correction to $P(11)_{\text{F}}$ is independent of statistics and thus the same result also holds for bosonic wave packets. The difference between bosons and fermions shows up in the interference phase between exchange ($1\to 2$) and direct ($1 \to 1$) trajectories of the particles in order $t^{4}$ of the tunneling amplitude~\cite{Note2}.

\prlsection{Partial interference in an extended collider}
Above, we demonstrated the apparent bunching of fermions ($\mathcal{Q}_{\text{Cl;F}} \leq 0$) in the weak tunneling limit ($t\ll 1$) and with suppressed interference. Next, we present the evaluation of $P(11)_{\text{F/B}}$ in the generic case, for both fermions (F) and bosons (B). We start by discussing the single-particle processes, where self-interference plays an important role in an extended collider. Let $A_{n}^{(1\rightarrow 1)}$ denote the single-particle amplitude of the process where the particle starts from $S_{1}$ and performs $n$ windings before reaching $D_1$. If we had full interference between different trajectories, the transmission probability would be $\mathcal{P}(1\rightarrow 1)_{\text{F/B}} = \big| \sum_{n=0}^{\infty} A_{n}^{(1\rightarrow 1)} \big|^{2}$,  for both fermions and bosons, since we consider only a single particle. With partial interference (due to wave packet size or decoherence~\cite{PhysRevB.106.245421}), trajectories with large winding number mismatch interfere less.   The probability $\mathcal{P}(1\rightarrow 1)_{\text{F/B}}$ with partial interference can be obtained by the substitution, $A_{n}^{(1\rightarrow 1)*}A_{m}^{(1\rightarrow 1)} \rightarrow f(|n-m|) A_{n}^{(1\rightarrow 1)*}A_{m}^{(1\rightarrow 1)} $  where $f(|n-m|)$ is an appropriate decay function that depends on the wave packet shape, Sec.~IV of SM~\cite{Note2}. For classical particles, there is no interference ($f(|n-m|)=\delta_{m,n}$) and hence $\mathcal{P}(1\rightarrow 1)_{\text{Cl}} =  \sum_{n=0}^{\infty} |A_{n}^{(1\rightarrow 1)} |^{2}$.

In practice, the sources emit wave packets with finite width which the extended collider breaks into a sequence of smaller wavelets, as depicted in Fig.~\ref{fig:extended_collider}a, giving rise to partial interference. (For simplicity, we set $\hbar=1$ and assume a linear dispersion relation so that the wave packets retain their shape and velocity.) With a scatterer of size $L$ (see  Fig.~\ref{fig:extended_collider}a), the particle with momentum $k$ gains an orbital phase $\varphi=kL$. Transmission (analogously for reflection) of an incoming wave packet from $(x_{1}^{(0)},t_{1}^{(0)})$ in $S_1$ to $(x_{1},t_{1})$ in $D_1$, is given as a convolution of the incoming wave packet $\tilde{\phi}_{s_{1}}$ and the transmission function, 
\begin{equation}\label{eq:transmitted_wavelet}
    \tilde{\phi}_{d_{1}}(X) = \int_{-\infty}^{\infty}dx \tilde{\phi}_{s_{1}}(x-x_{1}^{(0)}+vt_{1}^{(0)})
    \mathcal{T}(x-x_{1}+vt_{1}), 
\end{equation}
where $X=x_{1}-x_{1}^{(0)}-vt_{1}+vt_{1}^{(0)}$, $v$ is the propagation velocity and the transmission function $\mathcal{T}(x)$ is 
\begin{equation}\label{eq:transmission_coeff}
    \mathcal{T}(x) =  \rho_{1}\delta(x) +\tau_{1}\tau_{2}\rho_{2}\delta(x-L) 
    +\tau_{1}\tau_{2}\rho_{2}^{3}\delta(x-2L)+\dots 
     \end{equation}
Therefore, the transmitted wave packet is a superposition of an infinite number of wavelets separated by $L$.

The finite width of the wave packet and time delay in reaching the detector give rise to a suppression of interference in the transmission probability $P(1\rightarrow 1)_{\text{F/B}}$. For example, consider a particular case, where one wavelet performed $n$ windings and the other performed $m$ windings. By using Eqs.~\eqref{eq:transmitted_wavelet}--\eqref{eq:transmission_coeff}, one gets two wavelets, $\tau_{1}\tau_{2}\rho_{2}^{2n-1}\tilde{\phi}_{\text{s}_{1}}(X+nL)$ and $\tau_{1}\tau_{2}\rho_{2}^{2m-1}\tilde{\phi}_{\text{s}_{1}}(X+mL)$. Here, the  centers of the wavelets are separated by $|(n-m)L|$.
The probability amplitude is given by the superposition of an infinite number of wavelets of the mentioned forms. With increasing difference in the winding numbers $|n-m|$, the interference contribution to the transmission probability is reduced. For a model square wave packet (in momentum space) emitted by a dilutor QPC, the decay function for $n\neq m$ is given as $f(|n-m|)=\frac{i\ell}{L|m-n|}e^{-ik_{f}L|n-m|}(e^{-i\frac{L}{\ell} |m-n|}-1)$ where we have $\ell=\frac{v}{eV}$ with $V$ as the bias voltage between the  dilutor and the source edge, and $v$, $k_{f}$ denote the velocity  and the Fermi momentum~\footnote{While the parameter $\ell$ controls the decay function, the width of the wave packet depends on the smearing of its momentum space distribution.}.

Consider next a two-particle process, where each of the sources $S_1$ and $S_2$ simultaneously emits one particle. We now consider the case of partial interference between two wave packets and evaluate the probability $P(11)_{\text{F/B}}$ that each of the detectors registers one particle. There are two sets of processes which contribute to the two-particle probability amplitude: direct and exchange processes. While for fermions these processes sum up with opposite signs leading to enhanced antibunching, for bosons they add up with same signs. This is a direct manifestation of the  statistical exchange phase~\cite{doi:10.1142/5752}. More specifically, each exchange process introduces a factor $e^{i\pi\nu}$, with $\nu=0$ for bosons and $\nu=1$ for fermions. In the direct process the particle starts from $S_{1}$ ($S_{2}$) and goes to $D_{1}$ ($D_{2}$). The two-particle amplitude is $A_{\text{direct}} = \sum_{n,m} A_{n}^{(1\rightarrow1)}A_{m}^{(2\rightarrow2)} $. Direct processes involve an integer number of windings around the anti-dot; a full winding leads to an even number of exchanges and thus the amplitude comes with $+1$ phase for both  bosons and fermions, see Eq.~\eqref{eq:p11bf}. 
\begin{figure}[tbp!]
\centering
\includegraphics[width=\columnwidth, keepaspectratio]{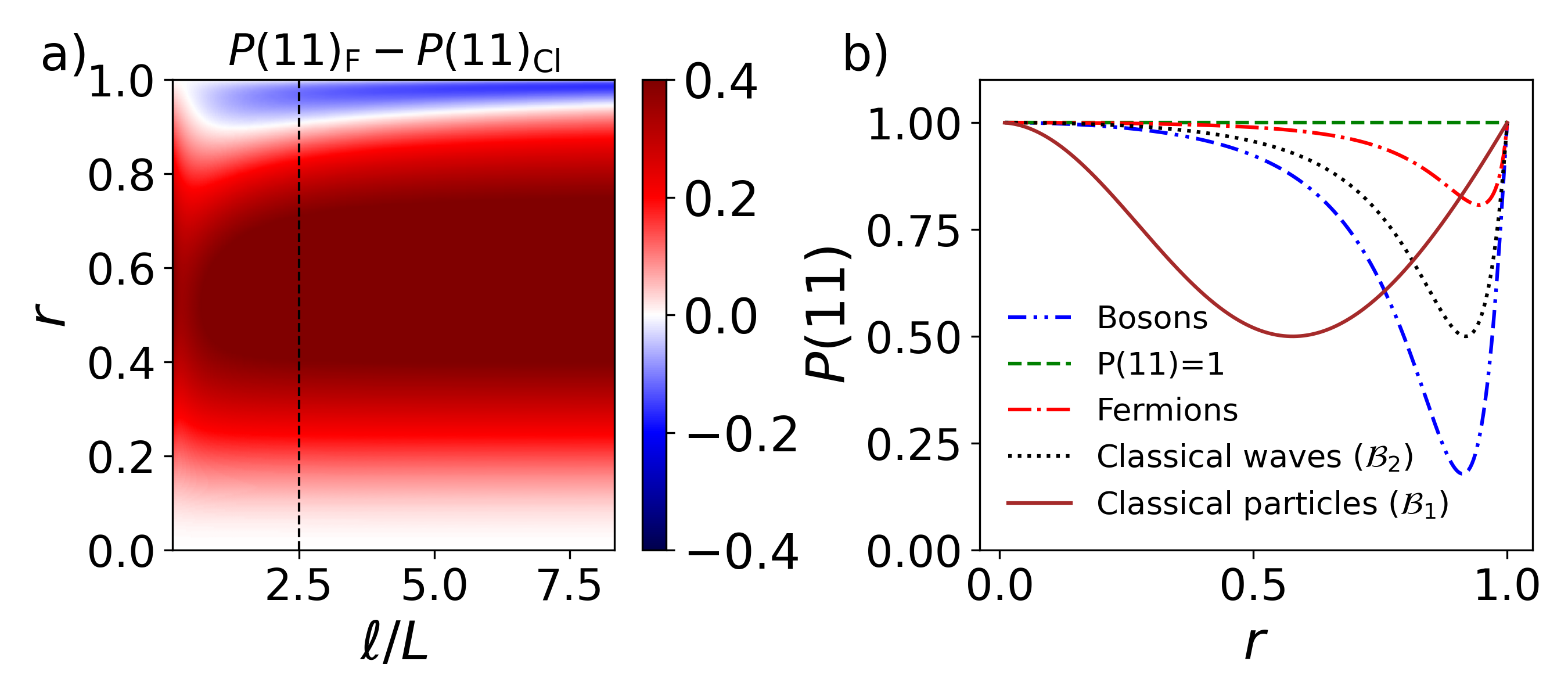}
\caption{Apparent and truly statistical measures of anti-bunching. \textbf{(a)}~Difference between the anti-bunching probabilities $P(11)_{\text{F}}$, Eq.~\eqref{eq:p11bf} of fermions  and  $P(11)_{\text{Cl}}$ of classical particles as functions of the wave packet width and the edge-QAD reflection amplitude $r$, Eq.~\eqref{eq:smatrix}. Use of the classical particles as a benchmark shows an apparent bunching of fermions (blue region). We take $k_f = 0$ and $\theta = 0$. \textbf{(b)}~Anti-bunching probabilities $P(11)$ for a model square wave packet with $\frac{\ell}{L}=2.5$, where $\ell=v/eV$~\cite{Note4} with $V$ as the bias voltage at the dilutor QPC and $v$ is the velocity, Sec. IV of SM~\cite{Note2}. Using $P(11)_{\text{Cl}}$ for classical particles as a  benchmark ($\mathcal{B}_{1}$) shows the apparent bunching of fermions near $r = 1$. The classical wave benchmark ($\mathcal{B}_{2}$) on the other hand reflects the true quantum statistics of fermions and bosons.}
  \label{fig:p11andg2}
  \end{figure}
In the exchange process the particle starts from $S_{1}$ ($S_{2}$) goes to $D_{2}$ ($D_{1}$) and there will be an odd number of exchanges leading to an extra phase, $(-1)^{\nu}$, Eq.~\eqref{eq:p11bf}. The  corresponding amplitude of the exchange processes is 
$A_{\text{exchange}} =   \sum_{n,m} A_{n+\frac{1}{2}}^{(1\rightarrow2)}A_{m+\frac{1}{2}}^{(2\rightarrow1)}$. The two-particle probability can be written as 
    \begin{equation}\label{eq:p11bf}
P(11)_{\text{F/B}} =|A_{\text{direct}} + (-1)^{\nu} A_{\text{exchange}}|^{2} 
\end{equation}
where the substitution, $A_{n}^{(i \rightarrow j)*}A_{m}^{(k \rightarrow j)} \rightarrow f({|n-m|}) A_{n}^{(i\rightarrow j)*}A_{m}^{(k \rightarrow j)} $ is to be used to account for interference suppression for the amplitudes to end in the same detector $D_j$ with $j=1,2$. With full interference, $f(n)=1$ for all $n$, we obtain $P(11)_{\text{F}} = 1$ but generically this is not the case.  
Partial interference of wave packets emitted from the sources leads to an apparent bunching of fermions for certain values of $r$, i.e. $\mathcal{Q}_{\text{Cl;F}}=P(11)_{\text{F}} - P(11)_{\text{Cl}}\leq 0$ (see Fig.~\ref{fig:p11andg2}a-b). 
Hence, we find that the benchmark, $\mathcal{B}_{1}=P(11)_{\text{Cl}}$ does not give the correct statistical information. In order to extract two-particle interference terms that reflect mutual statistics, one needs to subtract the contributions from single-particle interference terms. 
This can be achieved by defining a new benchmark $\mathcal{B}_{2}=\mathcal{P}(1\to1)_{s_{1}}\mathcal{P}(2\to2)_{s_{2}}+\mathcal{P}(1\to2)_{s_{1}}\mathcal{P}(2\to1)_{s_{2}}$, where the subscript denotes the active source.  
Since $\mathcal{B}_{2}$ consists only of single-particle probabilities, it is equivalent to the bunching probability of classical waves, Sec.~V of SM~\cite{Note2}; its value falls in between those of bosons and fermions, $P(11)_{\text{B}} < \mathcal{B}_{2} < P(11)_{\text{F}} $, Fig.~\ref{fig:p11andg2}b.

\prlsection{Current correlations}
Next, we show that the antibunching probability $P(11)_{\text{F/B}}$, as well as the benchmarks $\mathcal{B}_{1},\mathcal{B}_{2}$ can be accessed by measuring current correlations at the detectors. We begin by noting that the above analysis assumed no more than two particles are present simultaneously in the collider. From an experimental point of view, it is challenging to emit a single wave packet from a source. However, with recent developments in single electron emitters~\cite{PhysRevLett.101.166802,doi:10.1126/science.1141243,dubois_minimal-excitation_2013,doi:10.1126/science.284.5412.299,jullien_quantum_2014}, one can obtain a dilute stochastic beam of electrons. To model this, we assume that the sources emit Poisson distributed particles with an emission rate $\lambda$. The typical distance between emitted particles is $v / \lambda$, while the spatial size of the scattered particle (consisting of multiple wavelets) is of order $L$,  giving us the requirement $L \ll v/\lambda$ to not  have too many particles in the collider (see Sec.~VI of SM~\cite{Note2} for more detailed dependence of this estimate on the reflection amplitude $r$). Recall also that we obtain partial interference and can treat the collider as extended when the wave packet size is relatively small,   $\ell \lesssim L$. The width ($\ell$) of incoming wave packet is controlled by the bias voltage between the dilutor and the source edge~\cite{doi:10.1126/science.1141243,dubois_minimal-excitation_2013}. By tuning the bias voltage, one can achieve wave packets with widths comparable to the size ($L$) of the currently fabricated QADs which are typically of the order of a few hundreds of nanometers~\cite{PhysRevB.49.17456,PhysRevB.77.115328,PhysRevB.100.245130}.

We now account for the fact that the incoming beams represent  a mixed state consisting of a Poisson-distributed sequence of particles emitted from the sources (see Sec.~VII in SM~\cite{Note2}). Upon entering drain $i$, a particle passes an ideal detector at a fixed position $x_{i}$ which tracks the particles that ended up in that drain. The antibunching probability $P(11)_{\text{F/B}}$, Eq.~\eqref{eq:p11bf}, can then be expressed in terms of equal time cross-current correlations at the detectors $D_{1}$ and $D_{2}$, $ \langle I_{d_{1}} I_{d_{2}} \rangle \equiv \langle I_{d_{1}}(t) I_{d_{2}}(t) \rangle$, normalized by the average currents in the sources $\langle I_{s_{i}} \rangle=e\lambda$:
\begin{equation}\label{eq:cross_correlation}
P(11)_{\text{F/B}}  =  \frac{{\langle I_{d_{1}}I_{d_{2}} \rangle}_{s_{1}s_{2}}}{\langle I_{s_{1}} \rangle \langle I_{s_{2}} \rangle} 
 - \frac{{\langle I_{d_{1}}I_{d_{2}} \rangle}_{s_{1}} }{\langle I_{s_{1}} \rangle^2} 
 - \frac{{\langle I_{d_{1}}I_{d_{2}} \rangle}_{s_{2}} }{\langle I_{s_{2}} \rangle^2} 
  \,, 
\end{equation}  
where the subscript outside the angular brackets denotes the active source(s). For a chiral channel the current at the detector is given as $I_{d_{i}}(t_{i}) = evn_{d_{i}}(t)$, where $n_{d_{i}}$ is the density at the detector. Here, we have  subtracted the correlators with one source turned off because both of the detected particles can come from the same source, a possibility not accounted for in our calculation of $P(11)_{\text{F/B}}$, Eq.~\eqref{eq:p11bf}. Such cross-current correlators, Eq.~\eqref{eq:cross_correlation} have also been previously studied in the context of Hanbury-Brown and Twiss platforms made up of quantum Hall edge states to probe mutual statistics of the participating anyons~\cite{PhysRevLett.91.196803,zhang2023measuring}. Finally, we note that in the limit of a dilute beam, collisions of  three or more particles give subleading  corrections of order $\lambda L/v$ to Eq.~\eqref{eq:cross_correlation}.

Information on mutual statistics can be extracted by comparing Eq.~\eqref{eq:cross_correlation} with a benchmark. As discussed previously, choosing benchmark $\mathcal{B}_{1}=P(11)_{\text{Cl}}$ gives false statistical information, Fig.~\ref{fig:p11andg2} and hence, it is inappropriate to use. Experimentally, $P(11)_{\text{Cl}}$ can be obtained from Eq.~\eqref{eq:cross_correlation} 
in the limit of narrow wave packets such that there are no interference terms. The appropriate benchmark for extracting the mutual statistics is $\mathcal{B}_{2}$ which removes all self-interference contributions and is given in terms of currents as:
\begin{align}\label{eq:b2benchmark}
    \mathcal{B}_{2} = \frac{\langle I_{d_{1}} \rangle_{s_{1}}}{\langle I_{s_{1}} \rangle}\frac{\langle I_{d_{2}} \rangle_{s_{2}}}{\langle I_{s_{2}} \rangle} + \frac{\langle I_{d_{2}} \rangle_{s_{1}}}{\langle I_{s_{1}} \rangle}\frac{\langle I_{d_{1}} \rangle_{s_{2}}}{\langle I_{s_{2}} \rangle}
\end{align}
By subtracting $\mathcal{B}_{2}$, Eq.~\eqref{eq:b2benchmark} from $P(11)_{\text{F/B}}$, Eq.~\eqref{eq:cross_correlation} we are only left with the part of $P(11)_{\text{F/B}}$ that is sensitive to quantum statistics.

The two benchmarks $\mathcal{B}_{1}$ and $\mathcal{B}_{2}$ can also distinguish a point-like collider from an extended one. The presence of self-interfering trajectories leads to unequal benchmarks, $\mathcal{B}_{1}\neq\mathcal{B}_{2}$, while for point-like colliders, $\mathcal{B}_{1}=\mathcal{B}_{2}$ due to absence of any self-interference. Furthermore, the time-resolved density-density correlation function, $g^{(2)}$~\cite{loudon_quantum_2000} can distinguish point-like and extended colliders, Sec.~VIII of SM~\cite{Note2}. Additionally, the zero-frequency irreducible $g^{(2)}$~\cite{Note1} also gives information on mutual statistics, similar to our earlier discussions on cross-current correlations. Recent advances in single-electron emitters and quantum tomography~\cite{doi:10.1126/science.1141243,PhysRevLett.101.166802,dubois_minimal-excitation_2013,doi:10.1126/science.284.5412.299,jullien_quantum_2014} suggest that such time-resolved measurements are now feasible.

\prlsection{Conclusions} Our study demonstrates that pinpointing mutual statistics of quantum particles is subtle and requires analysis beyond idealized models. In a non-point-like collider considered here, the subtlety arises from multi-path interference. We analyzed a paradigmatic model of an extended collider, which offers an infinite number of possible trajectories for any single particle event, Fig.~\ref{fig:extended_collider}b, in contrast to a unique, well-defined trajectory in a point-like collider. To probe quantum statistics, we considered two possible benchmarks, $\mathcal{B}_{1}=P(11)_\text{Cl}$ and $\mathcal{B}_{2}=P(11)_\text{CW}$. Due to single-particle interference, $\mathcal{P}(1\to i)_{\text{CW}} = \mathcal{P}(1\to i)_{\text{Cl}}+(-1)^{i}\Delta$, $i=1,2$ and $\Delta$ encode the multi-path interference contribution~\cite{Note2}. Hence, for an extended collider,  the two benchmarks differ from each other, unlike in the case of  a point-like collider. The benchmark $\mathcal{B}_{2}$ removes the single-particle interference effects which may blur (in $\mathcal{B}_{1}$)  quantum statistics signatures in an extended collider. One may expect our general 
approach to hold beyond the class of chiral platforms, noting that studies of current correlation measurements and statistics have been extensively investigated in the case of nanowires and impurity scatterers, both in theory and experiment~\cite{Blanter_2000,PhysRevB.46.12485,PhysRevB.67.205408,PhysRevB.67.205408,PhysRevB.79.035121,PhysRevLett.104.026801,hofstetter2009cooper}. “Extended colliders” may arise due to geometric arrangement of impurities or due to inherent
 resonant-level spectra, facilitating the use of historic current correlation studies in  the modern context of quantum statistics effects. While the effect of electron-electron interaction demands further study, it is remarkable that one may construct colliders where the effect of interactions is small~\cite{zhang2023measuring}.

Extending our work to anyonic platforms is a most intriguing challenge, given the role of time-domain braiding~\cite{PhysRevLett.116.156802,Lee_2022,Han_2016} in auto- and cross-current correlations which is essentially a single source contribution. Unlike fermions where we eliminate single-source contributions
similar to Ref.~\cite{zhang2023measuring,zhang_fractional-statistics-induced_2025} for extracting mutual statistics, it is important to investigate the extent of single source contribution in the case of anyons. We also note that there is a wide range of platforms on which our analysis can be tested. Some examples: a collider made up of helical edges (edge modes of a time reversal invariant topological insulator); a collider coupled to a superconductor that, in addition to incoming particle–outgoing particle channels allows also for Andreev reflection channels (incoming particle—outgoing hole); collider made up of Abelian and non-Abelian anyon channels (including composite channels that allows for hole-like fractional quantum Hall phases and exotic neutralon excitations) and the realm of ultracold
atoms where key aspects of our work like interference pathways,
comparing bosons and fermions, collisions in a range of settings, and tunable interactions can be tested~\cite{PhysRevA.85.041601,chien_quantum_2015,PhysRevA.93.011606,PhysRevA.97.053614,PhysRevA.102.031302}.

\prlsection{Acknowledgements} Y.G. acknowledges support by
the DFG Grant MI 658/10-2. Y.G. and J.I.V.
acknowledge support by grant no 2022391 from the
United States - Israel Binational Science Foundation
(BSF), Jerusalem, Israel. 
Y. G. is supported by the Minerva
foundation. Y. G. is an incumbent of InfoSys chair at IISc. 
S.V. acknowledges the support of the National Science Foundation through Grant No. DMR-2004825.

\bibliography{refs}

\pagebreak
\widetext
\begin{center}
\textbf{\large Supplemental Materials: Quantum statistics and self-interference in extended colliders}
\end{center}
\setcounter{equation}{0}
\setcounter{figure}{0}
\setcounter{table}{0}
\setcounter{page}{1}
\makeatletter
\renewcommand{\theequation}{S\arabic{equation}}
\renewcommand{\thefigure}{S\arabic{figure}}

In this supplementary material, we show the details of the results discussed in the letter. We begin by giving the scattering formalism of our calculations in Sec.~\ref{sec:basicformalism}, which are independent of the shape of the wave packet that we choose and hence are applicable to all finite spatial width wave packets. 
In the following section, Sec.~\ref{sec:pointandextended} we demonstrate the differences between a point-like collider and the proposed extended collider. We show the differences between single-particle scattering and two-particle scattering in the two types of colliders. In the next section, Sec.~\ref{sec:toymodel}, we explain how an incorrect choice of benchmark can lead to a false statistical transmutation signal using a toy-model and show that its origin is related to the self-interfering single-particle trajectories (histories). In the section, Sec.~\ref{sec:dilutor}, we start by discussing wave packets and show that finite spatial width (of the wave packet) leads to partial interference. We also discuss a dilutor quantum point contact (QPC), which emits a dilute stochastic beam of particles (wave packets). In the following section, Sec.~\ref{sec:quantumstatistics} we define a benchmark for extracting the true mutual statistics of the colliding particles from the two-particle probabilities. The section, Sec.~\ref{sec:widthscatteredwave} gives the spatial width of the reflected and transmitted wave packet in the detector drains. Since we work with a stochastic beam of particles from the sources and our calculations are restricted to at most two particles simultaneously in the collider, we determine the appropriate emission rate $\lambda$ to have not more than two particles simultaneously in the collider. 
In the section, Sec.~\ref{sec:crosscurrentcorrelator} we calculate the cross-current correlator for a dilute stochastic beam of particles, and by defining a new benchmark in terms of the average currents, we extract the true statistics of the particles. In the final section, Sec.~\ref{sec:g2function}, we give a pedagogical definition of the $g^{(2)}-$function and then generalize it for a stochastic beam of particles denoted by $\bar{g}^{(2)}$. We demonstrate that the $\bar{g}^{(2)}-$function can be accessed experimentally by studying the auto-correlation of currents at a given detector. Finally, by defining an irreducible $\bar{g}^{(2)}$, referred to as $\bar{g}^{(2)}_{\text{Irr}}$, we show that this time-resolved function differentiates an extended collider from a point-like collider. Also, we demonstrate that the zero-frequency Fourier transform of $\bar{g}^{(2)}_{\text{Irr}}$ gives information on the mutual statistics of the particles.

\section{Scattering formalism}\label{sec:basicformalism}
In the setup, we have two chiral edges, with left moving particles on the lower edge and right moving particles on the upper edge. The chiral edges have linear dispersion, and the Hamiltonian of the system with right and left moving particles is given as,
\begin{align}
    \mathcal{H} = \int_{-\infty}^{\infty} dx \psi^{\dagger}(-i v \sigma_{z}\partial_{x})\psi(x)\,,
\end{align}
where $v$ is the magnitude of the velocity of the particles, and for simplicity we work in the units $\hbar=1$. The field operator $\psi(x)$ is given as,
\begin{align}
    \psi^{\dagger}(x) = \int_{-\infty}^{\infty} \frac{dk}{\sqrt{2\pi}} \left(\begin{array}{c}
         c_{k,R}^{\dagger}e^{-ikx}  \\
         c_{k,L}^{\dagger}e^{ikx}
    \end{array} \right)\,.
\end{align}
and $c_{k,R}^{\dagger}$ is the annihilation operator for the right movers, and $c_{k,L}^{\dagger}$ is the annihilation operator for the left moving particles and satisfies the anti-commutation relations, $\{c_{k,L},c^{\dagger}_{k',L}\}=2\pi\delta(k-k')=\{c_{k,R},c^{\dagger}_{k',R}\}$ and $\{c_{k,L},c^{\dagger}_{k',R}\}$=0. The collider, Fig.~\ref{fig:extended_collider}a consists of two sources $(S_{1}, S_{2})$ and two detectors $(D_{1},D_{2})$. At position $x=0$, we have a collider characterized by a $S-$matrix. On the left side (L) of the collider $(x<0)$, we have $c_{k,s_{1}}^{\dagger}$, the creation operator for the right moving particle coming from the source $S_{1}$ and $c_{k,d_{2}}^{\dagger}$, the creation operator for the left moving particle going into $D_{2}$. Correspondingly, on the right side (R) of the collider $(x>0)$, we have $c_{k,s_{2}}^{\dagger}$ for a left moving particle coming from $S_{2}$ and $c_{k,d_{1}}^{\dagger}$ for a right moving particle going into the detector $D_{1}$. Hamiltonian of the proposed setup, Fig.~\ref{fig:extended_collider}a, can be written as the sum of Hamiltonians on the left side of collider ($\mathcal{H}_{L}$) and right side of the collider ($\mathcal{H}_{R}$), i.e.,
\begin{align}
    \mathcal{H} &= \mathcal{H}_{L}+\mathcal{H}_{R}\,, 
\end{align}
where,
\begin{align}
    \mathcal{H}_{L} = \int_{-\infty}^{0} dx \psi_{L}^{\dagger}(x) (-i v \sigma_z) \psi_{L}(x) \ \ \text{and} \ \ \psi_{L}^{\dagger}(x) = \int_{-\infty}^{\infty} \frac{dk}{\sqrt{2\pi}} \left(\begin{array}{c}
         c^{\dagger}_{k,s_1}e^{-ikx} \\
         c^{\dagger}_{k,d_2}e^{ikx} 
    \end{array} \right) \,, \\ 
    \mathcal{H}_{R} = \int_{0}^{\infty} dx \psi_{R}^{\dagger}(x) (-i v \sigma_z) \psi_{R}(x) \ \ \text{and} \ \ \psi^{\dagger}_{R}(x) = \int_{-\infty}^{\infty} \frac{dk}{\sqrt{2\pi}} \left(\begin{array}{c}
         c^{\dagger}_{k,d_1}e^{-ikx} \\
         c^{\dagger}_{k,s_2}e^{ikx} 
    \end{array} \right)\,.
\end{align}
Once we have defined the creation/annihilation operators at the sources and detectors, these are related to each other by the unitary  $S-$matrix of the collider:  
\begin{align}\label{eq:extendedmatrix}
    \left(\begin{array}{c}
         c_{d_1,k}  \\
         c_{d_2,k} 
    \end{array} \right)= \left(\begin{array}{cc}
    \mathcal{T}_{k} & \mathcal{R}_{k}\\
    \mathcal{R}_{k} & \mathcal{T}_{k}
\end{array}\right)\left(\begin{array}{c}
         c_{s_1,k}  \\
         c_{s_2,k} 
    \end{array} \right)\,,
\end{align}
where the full expression for $\mathcal{T}_{k}$ and $\mathcal{R}_{k}$ are given later in  Sec.~\ref{sec:extendedcollider} (note that the $S-$matrix is momentum-dependent which is a result of non-point-like geometry of the collider, Fig.~\ref{fig:extended_collider}a).

In order to familiarize with the formalism, we start by understanding the single-particle physics and consider a specific example, where we create a wave packet at source $S_{1}$ and calculate the amplitude to find the particle in the detector, $D_{1}$. The operator $\Tilde{\phi}_{s_1}^{\dagger}$, creates a wave packet at $(x_{1}^{(0)},t_{1}^{(0)})$ originating from $S_{1}$ and is given as, 
\begin{align}
    \Tilde{\phi}^{\dagger}_{s_{1}}(x_{1}^{(0)},t_{1}^{(0)}) = \int_{-\infty}^{\infty} dk e^{ik(-x_{1}^{(0)}+vt_{1}^{(0)})} \phi^{*}(k) c_{k,s_{1}}^{\dagger} \,,
\end{align}
where $\phi(k)$ is the wave form in momentum space, normalized such that  $\int_{-\infty}^{\infty}dk |\phi(k)|^{2}=1$. The creation operator at position $(x_{1},t_{1})$ in the detector lead ($D_{1}$) is given as,
\begin{align}
    \phi_{d_{1}}^{\dagger}(x_{1},t_{1}) = \int_{-\infty}^{\infty} \frac{dk}{\sqrt{2\pi}} e^{ik(-x_{1}+vt_{1})} c_{k,d_{1}}^{\dagger} \,.
\end{align}
The amplitude for propagation to  $D_1$ is 
\begin{align}
    A(S_{1}\rightarrow D_{1}) =& \langle \Omega| \phi_{d_{1}}(x_{1},t_{1})\Tilde{\phi}_{s_{1}}^{\dagger}(x_{1}^{(0)},t_{1}^{(0)}) |\Omega\rangle\,, \\
    =& \int_{-\infty}^{\infty} \frac{dk}{\sqrt{2\pi}} \phi^{*}(k) \mathcal{T}_{k}e^{ik(x_{1}-x_{1}^{(0)}-vt_{1}+vt_{1}^{(0)})}\,,
\end{align}
where we have used Eq.~\eqref{eq:extendedmatrix} and the expectation value is taken with respect to a state $|\Omega\rangle$, which is a filled Fermi-sea at zero temperature. For calculating the probability from the amplitude, we integrate over the detection time $t_{1}$, from $t_{1}^{(0)}$ to $\infty$ inorder to ensure that the particle has reached the detector: 
\begin{align}
    \mathcal{P}(S_{1}\rightarrow D_{1}) = \int_{t_{1}^{(0)}}^{\infty} vdt_{1}|A(S_{1}\rightarrow D_{1})|^{2}\,.
\end{align}
Now, note that we have wave packets and because of causality, particle reaches the detector only after the time, $t_{1}^{(0)}$. The amplitude of finding the the particle at the detector for $t<t_{1}^{(0)}$ is zero and therefore we can effectively replace the lower limit of integration from $t_{1}^{(0)}$ to $-\infty$. Hence, we have,
\begin{align}
    \mathcal{P}(S_{1}\rightarrow D_{1}) = \int_{-\infty}^{\infty} vdt_{1}|A(S_{1}\rightarrow D_{1})|^{2} = \int_{-\infty}^{\infty} dk |\mathcal{T}_{k}|^{2}|\phi(k)|^{2}\,.
\end{align}
Similarly, we obtain the probability of being detected in $D_2$, 
\begin{align}
    \mathcal{P}(S_{1}\rightarrow D_{2}) = \int_{-\infty}^{\infty} dk |\mathcal{R}_{k}|^{2}|\phi(k)|^{2}\,.
\end{align}
Using the above scheme, we calculate the detection probabilities for point-like collider and extended collider in the following sections. For simplicity in the notations, we would like to denote the single-particle probability from the source to drain as $\mathcal{P}(i\to j)$ instead of $\mathcal{P}(S_{i}\to D_{j})$ and amplitudes as $A(i\to j)$ instead of $A(S_{i}\to D_{j})$.

\section{Antibunching probability $P(11)$ in point-like and  extended colliders}\label{sec:pointandextended}
In this section we are going to elaborate on the characteristic differences between a point-like and an extended collider. We begin by discussing  single-particle physics  and then   two-particle physics.
\subsection{Single-particle scattering in a point-like collider}
In a point-like collider the $S-$matrix is momentum independent and is given as,
\begin{align}\label{eq:pointcollidersmatrix}
    S=\left(\begin{array}{cc}
    \mathcal{T} & \mathcal{R}\\
    \mathcal{R} & \mathcal{T}
\end{array}\right)\,.
\end{align}
Once the wave packet goes through the collider, a part of every incoming wave packet will go to the lead with detector $D_{1}$ (transmission) and other part will go to lead with $D_{2}$ (reflection). Lets focus on the transmitted wave packet and assume $x_{1}$ be the location of the detector, $D_{1}$. Transmitted wave packet is related to the incoming one via the $S-$matrix, Eq.~\eqref{eq:pointcollidersmatrix}. Transmission amplitude is, 
\begin{align}
    A(1\rightarrow 1) = \langle \Omega| \phi_{d_1}(x_{1},t_{1}){\tilde{\phi}_{s_1}^{\dagger}(x_{1}^{(0)},t_{1}^{(0)})} |\Omega\rangle = \int_{-\infty}^{\infty} \frac{dk}{\sqrt{2\pi}} \mathcal{T} e^{ik(x_{1}-x_{1}^{(0)}-vt_{1}+vt_{1}^{(0)})}\phi^{*}(k)\,,
\end{align}
with $\phi^{\dagger}_{d_1}(x_{1},t_{1}) = \int_{-\infty}^{\infty} dk e^{ik(-x_{1}+vt_{1})}c^{\dagger}_{k,d_{1}}$ is the creation operator in the detector lead at $(x_{1},t_{1})$ and the expectation is with respect to the filled Fermi-sea, $|\Omega\rangle$. Since our $S-$matrix parameters in the point-like collider are not momentum dependent and therefore we can pull out the $\mathcal{T}$ from the integration. To obtain the detection probability, we integrate over all possible detection time, $t_{1}$,
\begin{align}
    \mathcal{P}(1\rightarrow 1) = \int_{t_{1}^{(0)}}^{\infty} vdt_{1} |A(1\rightarrow 1)|^{2} \,.
\end{align}
By using the argument of causality as in the previous section, Sec.~\ref{sec:basicformalism}, we can effectively replace the lower limit $t_{1}^{(0)}$ with $-\infty$. This implies, 
\begin{align}
    \mathcal{P}(1\rightarrow 1) &= \int_{-\infty}^{\infty} vdt_{1} |A(1\rightarrow 1)|^{2}\,, \\
    &= \int_{-\infty}^{\infty} \frac{vdt_{1}}{2\pi} |\mathcal{T}|^{2} \int_{-\infty}^{\infty} dk_{1} \int_{-\infty}^{\infty} dk_{2} e^{-ik_{1}(x_{1}-x_{1}^{(0)}-vt_{1}+vt_{1}^{(0)})} e^{ik_{2}(x_{1}-x_{1}^{(0)}-vt_{1}+vt_{1}^{(0)})} \phi(k_{1})\phi^{*}(k_{2}) \,.
\end{align}
By performing the time integral first we get $\delta(k_{1}-k_{2})$ and this can be simplified to,
\begin{align}
    \mathcal{P}(1\rightarrow 1) &= \int_{-\infty}^{\infty} dk |\phi(k)|^{2} |\mathcal{T}|^{2}\,, \\
    &= |\mathcal{T}|^{2}\,.
\end{align}
Here, our transmission amplitude is independent of momentum and therefore we can take it out of the integration. Hence, the probability of transmission is equal to $|\mathcal{T}|^{2}$ and similarly, one finds $\mathcal{P}(1\to 2)=|\mathcal{R}|^{2}$.

\subsection{Single-particle scattering in an extended collider}\label{sec:extendedcollider}
Extended collider is realized with the following effective $S-$matrix that we calculate using geometric series,
\begin{align}\label{eq:effectivesmatrix}
    S=\left(\begin{array}{cc}
    \mathcal{T}_k & \mathcal{R}_{k}\\
    \mathcal{R}_{k} & \mathcal{T}_{k}
\end{array}\right) = \left(\begin{array}{cc}
    \rho_{1}+\frac{\tau_{1}\tau_{2}\rho_{2}e^{ikL}}{1-\rho_{2}^{2}e^{ikL}} & \frac{\tau_{1}\tau_{2}e^{\frac{ikL}{2}}}{1-\rho_{2}^{2}e^{ikL}} \\
    \frac{\tau_{1}\tau_{2}e^{\frac{ikL}{2}}}{1-\rho_{2}^{2}e^{ikL}} & \rho_{1}+\frac{\tau_{1}\tau_{2}\rho_{2}e^{ikL}}{1-\rho_{2}^{2}e^{ikL}}
\end{array}\right) = \left(\begin{array}{cc}
    \frac{re^{i\theta}[1-e^{ikL-2i\theta}]}{1-r^{2}e^{ikL-i2\theta}} & -\frac{(1-r^{2})e^{i\frac{kL}{2}}}{1-r^{2}e^{ikL-i2\theta}} \\
    -\frac{(1-r^{2})e^{i\frac{kL}{2}}}{1-r^{2}e^{ikL-i2\theta}}& \frac{re^{i\theta}[1-e^{ikL-2i\theta}]}{1-r^{2}e^{ikL-i2\theta}}
\end{array}\right)\,.
\end{align}
where $\rho_{1},\rho_{2}$ are the amplitude to remain on the edge or in the quantum anti-dot (QAD) and $\tau_{1},\tau_{2}$ are the amplitude to hop in/out of the QAD at each junction in the extended collider, Eq.~(1). The transmission amplitude, $\mathcal{T}_{k}$ in the effective $S-$matrix, Eq.~\eqref{eq:effectivesmatrix} is obtained by the geometric series of all the possible processes, Fig.~\ref{fig:amplitudes}a,
\begin{align}
    \mathcal{T}_{k} = \rho_1 + \tau_{1}\tau_{2}\rho_{2}e^{ikL} + \tau_{1}\tau_{2}\rho_{2}^{3}e^{2ikL} + \cdots 
    = \rho_{1}+\frac{\tau_{1}\tau_{2}\rho_{2}e^{ikL}}{1-\rho_{2}^{2}e^{ikL}} \,.
\end{align}
As a result, for an extended collider, we get,
\begin{align}
     A(1\rightarrow 1) =& \langle\Omega|\phi_{d_{1}}(x_{1},t_{1})\tilde{\phi}^{\dagger}_{s_{1}}(x_{1}^{0},t_{1}^{0})|\Omega\rangle = \int_{-\infty}^{\infty} \frac{dk}{\sqrt{2\pi}} \mathcal{T}_{k} e^{ik(x_{1}-x_{1}^{(0)}-vt_{1}+vt_{1}^{(0)})}\phi(k)^{*}\,, \\
     =& \rho_{1} \int_{-\infty}^{\infty} \frac{dk}{\sqrt{2\pi}} e^{ik(x_{1}-x_{1}^{(0)}-vt_{1}+vt_{1}^{(0)})}\phi(k)^{*} \\
     & + \rho_{2}\tau_{1}\tau_{2} \int_{-\infty}^{\infty} \frac{dk}{\sqrt{2\pi}} e^{ik(x_{1}-x_{1}^{(0)}-vt_{1}+vt_{1}^{(0)}+kL)}\phi(k)^{*} + \cdots \,.\label{eq:wavelets} \\
     =& \int_{-\infty}^{\infty}dx \tilde{\phi}_{s_{1}}(x-x_{1}^{(0)}+vt_{1}^{(0)})
    \mathcal{T}(x-x_{1}+vt_{1})
\end{align}
where we have defined the following,
\begin{align}
   \tilde{\phi}_{s_{1}}(x) &= \int_{-\infty}^{\infty} dk e^{ik(x-x_{1}^{(0)}+vt_{1}^{(0)})}\phi^{*}(k) \\
   \mathcal{T}(x) &= \rho_1\delta(x) + \tau_{1}\tau_{2}\rho_{2}\delta(x-L) + \tau_{1}\tau_{2}\rho_{2}^{3}\delta(x-2L) + \cdots \,.    
\end{align}
 and one obtains Eq.~\eqref{eq:transmission_coeff} in the main text. This implies, we get an infinite number of wavelets in the transmission lead. Each of the wavelet have reduced amplitude and are separated by $L$, size of the detector, Fig.~\ref{fig:extended_collider}a. From here the probability is calculated as,
\begin{align}
    \mathcal{P}(1\rightarrow 1) &= \int_{t_{1}^{(0)}}^{\infty} vdt_{1} |A(1\rightarrow 1)|^{2} = \int_{-\infty}^{\infty} \frac{vdt_{1}}{2\pi} \Bigg| \int_{-\infty}^{\infty} dk \mathcal{T}_{k} e^{ik(x_{1}-x_{1}^{(0)}-vt_{1}+vt_{1}^{(0)})}\phi^{*}(k) \Bigg|^{2}\,, \label{eq:wp1simplip1to1}\\
    &=\int_{-\infty}^{\infty} dk |\mathcal{T}_{k}|^{2}|\phi(k)|^{2}\,. \label{eq:wp2simplip1to1}
\end{align}
In the limit where the spatial width of the wave packet is much smaller than the size of the QAD, $L$, then the amplitudes $A_{0}, \ A_{1}, \cdots$ in Eq.~\eqref{eq:wavelets} are well localized in position space separated by $L$. Therefore the interference terms like, $A_{n}^{*}A_{m}\rightarrow 0$ for $n\neq m$ and hence it reproduces the result for classical particles (subscript ``Cl" denotes the classical particles),
\begin{align}
    \mathcal{P}(1\rightarrow 1)_{\text{Cl}} = |\rho_{1}|^{2} + \frac{|\tau_{1}|^{2}|\tau_{2}|^{2}|\rho_{2}|^{2}}{1-|\rho_{2}|^{4}} =\frac{2r^{2}}{1+r^{2}}\,.
\end{align}

\subsubsection{Non-identical $S-$matrix at each channel-QAD junction}
We would like to finish this subsection by commenting on the case where the two junctions $S-$matrix are different. Let us consider the case where the upper junction is parameterized by $(r_{1},\theta_{1})$ and let the lower junction be parameterized by $(r_{2},\theta_{2})$, Eq.~\eqref{eq:smatrix}. Consider a single particle coming from the source $S_{1}$ and let us assume that the particle have momentum $k$. As a result of this, the effective matrix is modified in this case. Let us denote the new effective matrix with $S_{\text{Non-identical junction}}$,
\begin{align}\label{eq:nonidenticaljunctions}
    S_{\text{Non-identical junction}}=\left(\begin{array}{cc}
    \frac{e^{i\theta_{1}}[ r_{1}-r_{2}e^{i(kL-\theta_{1}-\theta_{2})} ]}{1-r_{1}r_{2}e^{i(kL-\theta_{1}-\theta_{2})}} & -\frac{\sqrt{(1-r_{1}^{2})(1-r_{2}^{2})}e^{ikL/2}}{1-r_{1}r_{2}e^{i(kL-\theta_{1}-\theta_{2})}} \\
    -\frac{\sqrt{(1-r_{1}^{2})(1-r_{2}^{2})}e^{ikL/2}}{1-r_{1}r_{2}e^{i(kL-\theta_{1}-\theta_{2})}} & \frac{e^{i\theta_{2}}[ r_{2}-r_{1}e^{i(kL-\theta_{1}-\theta_{2})} ]}{1-r_{1}r_{2}e^{i(kL-\theta_{1}-\theta_{2})}}
\end{array}\right)
\end{align}
Hence, we find that the effective $S-$matrix in the case of non-identical channel-QAD junctions also respects Time-reversal symmetry as in the identical case. From here, we compute the single-particle probability of transmission with a definite momentum $k$ is given as,
\begin{align}
\mathcal{P}_k(1\rightarrow1)=\mathcal{P}_k(2\rightarrow2)&=\big|\big[ S_{\text{Non-identical junction}} \big]_{11}\big|^{2}\,,\\
&=\frac{r_{1}^{2}+r_{2}^{2}-2r_{1}r_{2}\cos(kL-\theta_{1}-\theta_{2})}{1-2r_{1}r_{2}\cos(kL-\theta_{1}-\theta_{2})+r_{1}^{2}r_{2}^{2}}\,.
\end{align}
Similarly, the probability of reflection is given as,
\begin{align}
    \mathcal{P}_k(1\rightarrow 2) = \mathcal{P}_k(2\rightarrow 1) =&\big|\big[ S_{\text{Non-identical junction}} \big]_{12}\big|^{2}\,,\\
    =&\frac{(1-r_{1}^{2})(1-r_{2}^{2})}{1-2r_{1}r_{2}\cos(kL-\theta_1-\theta_2)+r_{1}^{2}r_{2}^{2}} \,.
\end{align}
In the case when we have identical junctions, we have resonant tunneling~\cite{gordanQM} in the system, $\mathcal{P}(1\to2)=\mathcal{P}(2\to1)=1$ for $kL=2\theta$, which is no more the case with different $S-$matrix at the junctions.

\subsection{Two-particle scattering in a point-like collider}
The two-particle probability can be computed in a similar way, we assume the detector $D_{1}$ is located at $x_{1}$ and detector $D_{2}$ is at $x_{2}$,
\begin{align}
    A(11)_{\text{B}/\text{F}} =& \langle\Omega|\phi_{d_{1}}(x_{1},t_{1})\phi_{d_{2}}(x_{2},t_{2})\tilde{\phi}_{s_{2}}^{\dagger}(x_{2}^{(0)},t_{2}^{(0)})\tilde{\phi}_{s_{1}}^{\dagger}(x_{1}^{(0)},t_{1}^{(0)})|\Omega\rangle\,, \\
    =& \langle\Omega|\phi_{d_{1}}(x_{1},t_{1})\tilde{\phi}_{s_{1}}^{\dagger}(x_{1}^{(0)},t_{1}^{(0)})|\Omega\rangle\langle\Omega|\phi_{d_{2}}(x_{2},t_{2})\tilde{\phi}_{s_{2}}^{\dagger}(x_{2}^{(0)},t_{2}^{(0)})|\Omega\rangle\,, \\
    & \pm \langle\Omega|\phi_{d_{1}}(x_{1},t_{1})\tilde{\phi}_{s_{2}}^{\dagger}(x_{2}^{(0)},t_{2}^{(0)})|\Omega\rangle\langle\Omega|\phi_{d_{2}}(x_{2},t_{2})\tilde{\phi}_{s_{1}}^{\dagger}(x_{1}^{(0)},t_{1}^{(0)})|\Omega\rangle \,,\\
     =& A_{\text{direct}} \pm A_{\text{exchange}}\,.
\end{align}
Now, inorder to get the probability for two-particle event, we integrate the detection time from $t_{0}=\min\{t_{1}^{(0)},t_{2}^{(0)}\}$ to $\infty$,
\begin{align}
    P(11)_{\text{B}/\text{F}} =& \int_{t_{0}}^{\infty}\int_{t_{0}}^{\infty} v^{2}dt_{1}dt_{2} |A_{\text{direct}} \pm A_{\text{exchange}}|^{2} \,.
\end{align}
Since, in the case of point-like collider, the transmission and reflection amplitude are independent of the momentum and by using the unitarity of the $S-$matrix, Eq.~\eqref{eq:pointcollidersmatrix} implies,
\begin{align}
    &S^{\dagger}S=\mathbb{I}\,, \\
    \Rightarrow \ &|\mathcal{R}|^{2} + |\mathcal{T}|^{2}=1 \ \ \text{and} \ \ \mathcal{R}^{*}\mathcal{T} + \mathcal{T}^{*}\mathcal{R} = 0\,. \label{eq:smatrixconstraints}
\end{align}
In the probability, $P(11)_{\text{B}/\text{F}}$, there are terms proportional to $\mathcal{T}^{*2}\mathcal{R}^{2}$ and then by using Eq.~\eqref{eq:smatrixconstraints}, we get, $\mathcal{T}^{*2}\mathcal{R}^{2}=-|\mathcal{R}|^{2}|\mathcal{T}|^{2}$ and hence we obtain,
\begin{align}
    P(11)_{\text{B}/\text{F}} = |\mathcal{R}|^{4} + |\mathcal{T}|^{4} \mp 2|J|^{2}|\mathcal{R}|^{2}|\mathcal{T}|^{2}\,, \ \ \text{where} \ \ J = \int_{-\infty}^{\infty} dk e^{ik( x_{2}^{(0)} + vt_{2}^{(0)} + x_{1}^{(0)} - vt_{2}^{(0)} )} |\phi(k)|^{2}\,.
\end{align}
Function $J$ is the overlap function \cite{Blanter_2000}, when the particles arrive at the collider at the same time, i.e, $x_{2}^{(0)} + vt_{2}^{(0)} + x_{1}^{(0)} - vt_{2}^{(0)}=0$ then, $J=1$. In the case with fermions arriving simultaneously in the collider, by Pauli's principle we expect them to leave for different detectors and consistently we also obtain $P(11)_{\text{F}} = 1$.

\subsection{Two-particle scattering in an extended collider}
Characteristic feature of an extended collider is the momentum dependent transmission and reflection coefficients in the effective $S-$matrix of the collider, Eq.~\eqref{eq:effectivesmatrix}. As a result, we get a train of wavelets in the drains, Eq.~\eqref{eq:wavelets}.  With detector $D_{1}$ at $x_{1}$ and $D_{2}$ at $x_{2}$, the amplitudes for direct and exchange processes are given as follows,
\begin{align}
    A_{\text{direct}} & =\langle\Omega|\phi_{d_{1}}(x_{1},t_{1})\tilde{\phi}_{s_{1}}^{\dagger}(x_{1}^{(0)},t_{1}^{(0)})|\Omega\rangle\langle\Omega|\phi_{d_{2}}(x_{2},t_{2})\tilde{\phi}_{s_{2}}^{\dagger}(x_{2}^{(0)},t_{2}^{(0)})|\Omega\rangle\,,\\
    & =\mathcal{I}_{1}(x_{1},t_{1};x_{1}^{(0)},t_{1}^{(0)})\mathcal{I}_{2}(x_{2},t_{2};x_{2}^{(0)},t_{2}^{(0)})\,, \label{eq:directamp11} \\
    A_{\text{exchange}} &= \langle\Omega|\phi_{d_{1}}(x_{1},t_{1})\tilde{\phi}_{s_{2}}^{\dagger}(x_{2}^{(0)},t_{2}^{(0)})|\Omega\rangle\langle\Omega|\phi_{d_{2}}(x_{2},t_{2})\tilde{\phi}_{s_{1}}^{\dagger}(x_{1}^{(0)},t_{1}^{(0)})|\Omega\rangle \,,\\
    &= \mathcal{J}_{2}(x_{1},t_{1};x_{2}^{(0)},t_{2}^{(0)})\mathcal{J}_{1}(x_{2},t_{2};x_{1}^{(0)},t_{1}^{(0)})\,, \label{eq:exchamp11}
\end{align}
where,
\begin{align}
\mathcal{I}_{1}(x_{1},t_{1};x_{1}^{(0)},t_{1}^{(0)}) &= \int_{-\infty}^{\infty} \frac{dk_{3}}{\sqrt{2\pi}}e^{ik_{1}(x_{1}-vt_{1}-x_{1}^{(0)}+vt_{1}^{(0)})}\phi^{*}(k_{1})\mathcal{T}_{k_{1}}\,, \\
\mathcal{I}_{2}(x_{2},t_{2};x_{2}^{(0)},t_{2}^{(0)}) &=  \int_{-\infty}^{\infty} \frac{dk_{2}}{\sqrt{2\pi}}e^{ik_{2}(-x_{2}-vt_{2}+x_{2}^{(0)}+t_{2}^{(0)})}\phi^{*}(k_{2})\mathcal{T}_{k_{2}} \,, \\
    \mathcal{J}_{1}(x_{2},t_{2};x_{1}^{(0)},t_{1}^{(0)}) &= \int_{-\infty}^{\infty} \frac{dk_{1}}{\sqrt{2\pi}}e^{ik_{1}(-x_{2}-vt_{2}-x_{1}^{(0)}+t_{1}^{(0)})}\phi^{*}(k_{1})\mathcal{R}_{k_{1}} \,, \\
    \mathcal{J}_{2}(x_{1},t_{1};x_{2}^{(0)},t_{2}^{(0)}) &= \int_{-\infty}^{\infty} \frac{dk_{2}}{\sqrt{2\pi}}e^{ik_{2}(x_{1}-vt_{1}+x_{2}^{(0)}+t_{2}^{(0)})}\phi^{*}(k_{2})\mathcal{R}_{k_{2}} \,.
\end{align}
The probability in this case is,
\begin{align}\label{eq:p11wavepackets}
    P(11)_{\text{B}/\text{F}} =& \int_{-\infty}^{\infty}\int_{-\infty}^{\infty} v^{2}dt_{1}dt_{2}|\mathcal{I}_{1}(x_{1},t_{1};x_{1}^{(0)},t_{1}^{(0)})\mathcal{I}_{2}(x_{2},t_{2};x_{2}^{(0)},t_{2}^{(0)}) \pm  \mathcal{J}_{2}(x_{1},t_{1};x_{2}^{(0)},t_{2}^{(0)})\mathcal{J}_{1}(x_{2},t_{2};x_{1}^{(0)},t_{1}^{(0)})|^{2}\,.
\end{align}
where, we have the lower limit by of the integration as $-\infty$ by using the arguments as in section Sec.~\ref{sec:basicformalism}. Further, we get,
\begin{align}
     P(11)_{\text{B}/\text{F}} =& \int_{-\infty}^{\infty}vdt_{1}|\mathcal{I}_{1}(x_{1},t_{1};x_{1}^{(0)},t_{1}^{(0)})|^{2}\int_{-\infty}^{\infty} vdt_{2} |\mathcal{I}_{2}(x_{2},t_{2};x_{1}^{(0)},t_{1}^{(0)})|^{2}\,, \\
     & + \int_{-\infty}^{\infty}vdt_{1}|\mathcal{J}_{2}(x_{1},t_{1};x_{2}^{(0)},t_{2}^{(0)})|^{2}\int_{-\infty}^{\infty} vdt_{2}|\mathcal{J}_{1}(x_{2},t_{2};x_{1}^{(0)},t_{1}^{(0)})|^{2} \,,\\
     & \pm 2\text{Re}\big[\chi(x_{1}^{(0)},t_{1}^{(0)};x_{2}^{(0)},t_{2}^{(0)})\big]\,,
\end{align}
where,
\begin{align}
\chi(x_{1}^{(0)},t_{1}^{(0)};x_{2}^{(0)},t_{2}^{(0)}) =  \iint_{t_{0}}^{\infty} v^{2}dt_{1}dt_{2} \mathcal{I}_{1}(x_{1},t_{1};x_{1}^{(0)},t_{1}^{(0)})\mathcal{I}_{2}(x_{2},t_{2};x_{2}^{(0)},t_{2}^{(0)}) \mathcal{J}_{2}^{*}(x_{1},t_{1};x_{2}^{(0)},t_{2}^{(0)})\mathcal{J}_{1}^{*}(x_{2},t_{2};x_{1}^{(0)},t_{1}^{(0)})\,.
\end{align}
This can be simplified further and we obtain,
\begin{align}
\chi(x_{1}^{(0)},t_{1}^{(0)};x_{2}^{(0)},t_{2}^{(0)}) =& \int_{-\infty}^{\infty} dk|\phi(k)|^{2}\mathcal{R}_{k}^{*}\mathcal{T}_{k}e^{ik\big[-x_{2}^{(0)}-vt_{2}^{(0)}-x_{1}^{(0)}+vt_{1}^{(0)}\big]}\\
 & \quad\times\int_{-\infty}^{\infty} dk'|\phi(k')|^{2}\mathcal{R}_{k'}^{*}\mathcal{T}_{k'}e^{-ik'\big[-x_{2}^{(0)}-vt_{2}^{(0)}-x_{1}^{(0)}+vt_{1}^{(0)}\big]}\,.
\end{align}
Similarly, the first two term in the expression for $P(11)_{\text{B/F}}$ can be simplified as follows,
\begin{align}
    \int_{-\infty}^{\infty}vdt_{1}|\mathcal{I}_{1}(x_{1},t_{1};x_{1}^{(0)},t_{1}^{(0)})|^{2}=&\int_{-\infty}^{\infty}\frac{vdt_{1}}{2\pi}\int_{-\infty}^{\infty} dk_{1}e^{ik_{1}(x_{1}-vt_{1}-x_{1}^{(0)}+vt_{1}^{(0)})}\phi^{*}(k_{1})\mathcal{T}_{k_{1}} \\
    &\times\int_{-\infty}^{\infty} dk_{1}'e^{-ik_{1}'(x_{1}-vt_{1}-x_{1}^{(0)}+vt_{1}^{(0)})}\phi(k_{1}')\mathcal{T}^{*}_{k_{1}'} \,,\\
    =&\int_{-\infty}^{\infty} dk_{1}|\phi(k_{1})|^{2}|\mathcal{T}_{k_{1}}|^{2}\,.
\end{align}
Here, we used the fact that the integral over time give rise to $\delta(k_{1}-k_{1}')$. By similar set of calculations, we get, $\int_{-\infty}^{\infty} vdt_{2}|\mathcal{J}_{1}(x_{2},t_{2};x_{1}^{(0)},t_{1}^{(0)})|^{2}=\int_{-\infty}^{\infty} dk_{1}|\phi(k_{1})|^{2}|\mathcal{R}_{k_{1}}|^{2}$. Therefore, the probability $P(11)_{\text{B/F}}$ can be written as, 
\begin{align}
    P(11)_{\text{B}/\text{F}} =& \Big( \int_{-\infty}^{\infty} dk |\mathcal{T}_{k}|^{2}|\phi(k)|^{2} \Big)^{2} + \Big( \int_{-\infty}^{\infty} dk |\mathcal{R}_{k}|^{2}|\phi(k)|^{2} \Big)^{2} \,,\\
    & \pm \chi(x_{1}^{(0)},t_{1}^{(0)};x_{2}^{(0)},t_{2}^{(0)})\,.
\end{align}
We would like to understand the interference term in the probability, $P(11)_{\text{B}/\text{F}}$. The interference term is given as,
\begin{align}
\chi(x_{1}^{(0)},t_{1}^{(0)};x_{2}^{(0)},t_{2}^{(0)}) =& \int_{-\infty}^{\infty} dk|\phi(k)|^{2}\mathcal{R}_{k}^{*}\mathcal{T}_{k}e^{ik\big[-x_{2}^{(0)}-vt_{2}^{(0)}-x_{1}^{(0)}+vt_{1}^{(0)}\big]}\\
 & \quad\times\int_{-\infty}^{\infty} dk'|\phi(k')|^{2}\mathcal{R}_{k'}^{*}\mathcal{T}_{k'}e^{-ik'\big[-x_{2}^{(0)}-vt_{2}^{(0)}-x_{1}^{(0)}+vt_{1}^{(0)}\big]}\,, \\
  =& \int_{-\infty}^{\infty} dk|\phi(k)|^{2}\mathcal{R}_{k}^{*}\mathcal{T}_{k}e^{ikx}\int_{-\infty}^{\infty} dk'|\phi(k')|^{2}\mathcal{R}_{k'}^{*}\mathcal{T}_{k'}e^{-ik'x} \,,\\
  =& -\Big|\int_{-\infty}^{\infty} dk|\phi(k)|^{2}\mathcal{R}_{k}^{*}\mathcal{T}_{k}e^{ikx}\Big|^{2}\leq 0 \,.\label{eq:interfterm}
\end{align}
where $x=x_{2}^{(0)}+vt_{2}^{(0)}+x_{1}^{(0)}-vt_{1}^{(0)}$. Hence, we obtain that $\chi$ is a real function and is always negative. This tells us that the fermions always have the tendency to anti-bunch with respect to bosons, i.e., $P(11)_{\text{F}}\geq P(11)_{\text{B}}$. Since the sources are located are equal distance from the collider but on the opposite sides, Fig.~\ref{fig:extended_collider}a, then for simultaneously arriving wave packets ($t_{1}^{(0)}=t_{2}^{(0)}$) and we have $x=0$ and we obtain $P(11)_{\text{F}}$ as shown in Fig.~\ref{fig:p11andg2}a,b. We end this subsection by evaluating the probability of the event $P(20)$ of receiving both the particles in the same detector, $D_{1}$ (or $D_{2}$). By using Eq.~\eqref{eq:directamp11} and Eq.~\eqref{eq:exchamp11}, we get,
\begin{align}
    P(20)_{\text{B/F}} = & \frac{1}{2} \iint_{t_{0}}^{\infty} v^{2}dt_{1}dt_{2} |\mathcal{I}_{1}(x_{1},t_{1};x_{1}^{(0)},t_{1}^{(0)})\mathcal{J}_{2}(x_{2},t_{2};x_{2}^{(0)},t_{2}^{(0)}) \pm  \mathcal{I}_{1}(x_{2},t_{2};x_{1}^{(0)},t_{1}^{(0)})\mathcal{J}_{2}(x_{1},t_{1};x_{2}^{(0)},t_{2}^{(0)})|^{2}\,.
\end{align}
The factor of $1/2$ comes from the fact that we have indistinguishable particles and we are integrating over the times $t_{1}$ and $t_{2}$ and both particles end up in the same detector. The probability can be written as,
\begin{align}
    P(20)_{\text{B/F}}=& \int_{-\infty}^{\infty}vdt_{1}|\mathcal{I}_{1}(x_{1},t_{1};x_{1}^{(0)},t_{1}^{(0)})|^{2}\int_{-\infty}^{\infty} vdt_{2} |\mathcal{J}_{2}(x_{2},t_{2};x_{1}^{(0)},t_{1}^{(0)})|^{2}\,, \\
     & \pm \text{Re}\big[\mathcal{G}(x_{1}^{(0)},t_{1}^{(0)};x_{2}^{(0)},t_{2}^{(0)})\big]=P(02)_{\text{B/F}}\,,
\end{align}
where,
\begin{align}
    \mathcal{G}(x_{1}^{(0)},t_{1}^{(0)};x_{2}^{(0)},t_{2}^{(0)}) =& \iint_{t_{0}}^{\infty} v^{2}dt_{1}dt_{2} \mathcal{I}_{1}(x_{1},t_{1};x_{1}^{(0)},t_{1}^{(0)})\mathcal{J}_{2}(x_{2},t_{2};x_{2}^{(0)},t_{2}^{(0)}) \mathcal{J}_{2}^{*}(x_{1},t_{1};x_{2}^{(0)},t_{2}^{(0)})\mathcal{I}_{1}^{*}(x_{2},t_{2};x_{1}^{(0)},t_{1}^{(0)})\,,\\
    =-& \chi(x_{1}^{(0)},t_{1}^{(0)};x_{2}^{(0)},t_{2}^{(0)}) \,.
\end{align}
In the case of classical waves (denoted with the subscript, ``CW"), we do not have the interference terms between direct and exchange processes and the probability with two active sources of waves is simply given as,
\begin{align}
    P(11)_{\text{CW}} =& \int_{-\infty}^{\infty}vdt_{1}|\mathcal{I}_{1}(x_{1},t_{1};x_{1}^{(0)},t_{1}^{(0)})|^{2}\int_{-\infty}^{\infty} vdt_{2} |\mathcal{I}_{2}(x_{2},t_{2};x_{1}^{(0)},t_{1}^{(0)})|^{2}\,, \\
     & + \int_{-\infty}^{\infty}vdt_{1}|\mathcal{J}_{2}(x_{1},t_{1};x_{2}^{(0)},t_{2}^{(0)})|^{2}\int_{-\infty}^{\infty} vdt_{2}|\mathcal{J}_{1}(x_{2},t_{2};x_{1}^{(0)},t_{1}^{(0)})|^{2} \,,\\
     =& \Big( \int_{-\infty}^{\infty} dk |\mathcal{T}_{k}|^{2}|\phi(k)|^{2} \Big)^{2} + \Big( \int_{-\infty}^{\infty} dk |\mathcal{R}_{k}|^{2}|\phi(k)|^{2} \Big)^{2} \,,\\
     =&\mathcal{P}(1\to1)^{2} + \mathcal{P}(1\to2)^{2}\,,\\
     P(20)_{\text{CW}} =& \int_{-\infty}^{\infty}vdt_{1}|\mathcal{I}_{1}(x_{1},t_{1};x_{1}^{(0)},t_{1}^{(0)})|^{2}\int_{-\infty}^{\infty} vdt_{2} |\mathcal{J}_{2}(x_{2},t_{2};x_{1}^{(0)},t_{1}^{(0)})|^{2}\,, \\
     =& \Big( \int_{-\infty}^{\infty} dk |\mathcal{T}_{k}|^{2}|\phi(k)|^{2} \Big)\Big( \int_{-\infty}^{\infty} dk |\mathcal{R}_{k}|^{2}|\phi(k)|^{2} \Big) \,,\\
     =& \mathcal{P}(1\to1)\mathcal{P}(1\to2)= P(02)_{\text{CW}}\,.
\end{align}

\section{Toy model for apparent bunching of fermions}\label{sec:toymodel}
\begin{figure}[t]
    \centering
    \includegraphics[width=16cm]{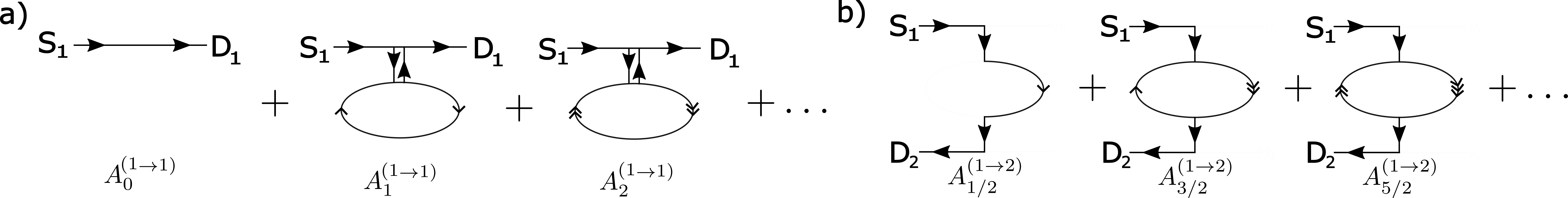}
    \caption{(a) Trajectories of the particle from the source $S_{1}$ to $D_{1}$. The amplitude for the process $A(1\rightarrow 1)$ is a sum of amplitudes of each of the diagram. (b) Trajectories of the particle from the source $S_{1}$ to $D_{2}$. The amplitude for the process $A(1\rightarrow 2)$ is a sum of amplitudes of each of the diagram. Amplitude of each of the trajectory can be written by using the junction $S-$matrix, Eq.~(1).}
    \label{fig:amplitudes}
\end{figure}

In this section we show an apparent bunching of fermions which corresponds to the fact that the antibunching probability of fermions, $P(11)_{\text{F}}$ recedes that of classical particles, $P(11)_{\text{Cl}}$. In other words, we have the inequality, $P(11)_{\text{F}}\leq P(11)_{\text{Cl}}$ for certain values of the tunneling amplitude.  
This illustrates that by choosing the benchmark $\mathcal{B}_{1}=P(11)_{\text{Cl}}$ doesn't not reflect the true statistical behavior of fermions/bosons. 
(In the later part of the supplemental material, Sec.~\ref{sec:quantumstatistics}, we define another  benchmark $\mathcal{B}_{2}$ which will successfully determine the underlying statistics.) Hence, the apparent statistical transmutation~\cite{PhysRevB.99.045430} is a result of an incorrect choice of benchmark and is not a true signature of quantum statistics. Consider a single particle coming in from source, $S_1$. Then, it can end up in either of the two detectors, $D_{1}$ to $D_{2}$. The diagrams, Fig.~\ref{fig:amplitudes}, shows the different trajectories for the particle to go from $S_{1}$ to $D_{1}$ or $D_{2}$. The amplitude for different processes can be written as,
\begin{equation}A_{0}^{(1\rightarrow 1)}=\rho_{1},\:A_{n\neq0}^{(1\rightarrow 1)}=\tau_{1}\tau_{2}\rho_{2}^{2n-1}e^{in\varphi}, \ \ \text{and} \ \ A_{n+\frac{1}{2}}^{(1\rightarrow 2)} = \tau_{1}\tau_{2}\rho_{2}^{2n}e^{i(n+\frac{1}{2})\varphi} \,.
\end{equation}
where the subscript $n$ refers to the winding number in the corresponding amplitude. Note that, the amplitude $A_{n}^{(1\rightarrow1)}$ comes with a phase $e^{i n \varphi}$, this is the orbital phase (equivalent to $kL$ in the section, Sec.~\ref{sec:extendedcollider}) acquired by the particle as it goes around the QAD $n$ number of times, Fig.~\ref{fig:amplitudes}a. Similarly, when the particle goes from $S_{1}$ to $D_{2}$, it will acquire a phase, $e^{i(n+\frac{1}{2})\varphi}$, Fig.~\ref{fig:amplitudes}b. With the single-particle amplitude written, single-particle probability in general can be written as follows,
\begin{align}
\mathcal{P}(1 & \rightarrow1)=\sum_{n,m=0}^{\infty}A_{n}^{(1\rightarrow 1)*}A_{m}^{(1\rightarrow1)}f(|n-m|)\,,
\end{align}
where the function $f(|n-m|)$ is the decay function for $m\neq n$ and determines the contribution of different interference terms ($A_{n}^{(1\rightarrow 1)*}A_{m}^{(1\rightarrow1)}$ when $m\neq n$) in the final probability and when $m=n$, we have $f(0)=1$. For classical particles, there are no interference between any of
the processes with different winding number and hence we set, $f(|n-m|)=0$ (for $m\neq n$) and we obtain,
\begin{align}
\mathcal{P}(1\rightarrow1)_{\text{Cl}} & =\sum_{n=0}^{\infty}A_{n}^{(1\rightarrow 1)*}A_{n}^{(1\rightarrow1)}=\frac{2r^{2}}{1+r^{2}}\,.
\end{align}
In the case of plane waves we have interference among all the processes and for fermions and bosons we obtain by setting $f(|n-m|)=1$, 
\begin{equation}
\mathcal{P}(1\rightarrow1)_{\text{F/B}}= \Big| \frac{re^{i\theta}[1-e^{i\varphi-2i\theta}]}{1-r^{2}e^{i\varphi-i2\theta}} \Big|^{2} = \frac{2r^{2}(1-\cos\alpha)}{1+r^{4}-2r^{2}\cos\alpha} \,,
\end{equation}
for all values of the tunneling amplitude, $r$ and $\alpha=\varphi-2\theta$ is the total geometric phase the particle gains as it goes around the QAD. Additionally, at $\varphi = 2\theta$ ($\alpha = 0$), we have resonance in system, which corresponds to the fact that $\mathcal{P}(1\to1)=0$~\cite{gordanQM}. For better clarity in understanding the system, we would like to introduce
the following notations,
\begin{align}
\mathcal{P}(1\rightarrow1) & =\sum_{n,m=0}^{\infty}A_{n}^{(1\rightarrow 1)*}A_{m}^{(1\rightarrow1)}f(|n-m|)\,,\\
 & =\mathcal{P}(1\rightarrow1)_{\text{Cl}}-\Delta\,,\\ \label{eq:p1to1atreso}
\text{where},\:\Delta & =-\sum_{n\neq m}^{\infty}A_{n}^{(1\rightarrow 1)*}A_{m}^{(1\rightarrow1)}f(|n-m|)\,.
\end{align}
This implies that for fermions and bosons with full interference, we have
$\mathcal{P}(1\rightarrow1)_{\text{F/B}}=0=\mathcal{P}(1\rightarrow1)_{\text{Cl}}-\Delta$ at resonance
and hence,
\begin{align}\label{eq:deltainfty}
\Delta= & \mathcal{P}(1\rightarrow1)_{\text{Cl}}=\frac{2r^{2}}{1+r^{2}} \ \ \text{(full interference and at resonance, $\alpha=0$)}\,,
\end{align}
Next, we would like to look at the case where we
have partial interference in the system. This corresponds to an intermediate case
where the contribution from the interference terms ($A_{n}^{(1\rightarrow 1)*}A_{m}^{(1\rightarrow1)}$ for $n\neq m$) is suppressed. Consider the simplest partial interference case, where we allow interference
only among processes where the winding number differs by 1, i.e. we
choose $f(|n-m|)$ as,
\begin{equation}\label{eq:decayfunction}
f(|n-m|)=c_{1}(\delta_{n,m-1}+\delta_{n,m+1}) \,,
\end{equation}
with $0<c_{1}\ll$1 is the suppression factor, and the suppresses the interference of the processes where have winding number differs by 1. We neglect all the interference processes where the winding number differs by more than 1. 
This implies,
\begin{align}
\mathcal{P}(1\rightarrow1)_{\text{F/B}} & =\sum_{n=0}^{\infty}A_{n}^{(1\rightarrow 1)*}A_{n}^{(1\rightarrow1)}+c_{1}\sum_{n=0}^{\infty}A_{n}^{(1\rightarrow 1)*}A_{n+1}^{(1\rightarrow1)}+c_{1}\sum_{n=0}^{\infty}A_{n+1}^{(1\rightarrow 1)*}A_{n}^{(1\rightarrow1)}\,,\\ \label{eq:p1to1decoherence}
 & =\mathcal{P}(1\rightarrow1)_{\text{Cl}}- \Delta_{1}\,,  \\
\mathcal{P}(1\rightarrow 1)_{\text{Cl}} &= \frac{2r^{2}}{1+r^2}=\frac{2(1-t^{2})}{2-t^{2}} \ \ \text{and} \  \ \Delta_{1} =\frac{2c_{1}r^{2}(1-r^{2})\cos2\alpha}{1+r^{2}}=\frac{2c_{1}(1-t^{2})t^{2}\cos2\alpha}{2-t^{2}} \,,
\end{align}
with $\alpha=\frac{\varphi}{2}-\theta$ and $t^2=1-r^2$. Similar set of calculation yields,
\begin{align}
    \mathcal{P}(1\rightarrow 2)_{\text{F/B}} &= \mathcal{P}(1\rightarrow 2)_{\text{Cl}}+ \Delta_{1} \,.\\
    \mathcal{P}(1\rightarrow 2)_{\text{Cl}} &= \frac{1-r^{2}}{1+r^{2}}=\frac{t^{2}}{2-t^{2}}\,. 
\end{align}
The two-particle probability, $P(11)$, where we receive
one particle at each of the detector is given as,
\begin{align}
P(11)_{\text{B/F}} & =|A_{\text{direct}}|^{2}+|A_{\text{exchange}}|^{2}\pm2\text{Re}[A_{\text{direct}}^{*}A_{\text{exchange}}]\,.
\end{align}
Here, $A_{\text{direct}}$ corresponds to the processes where the particle from $S_{1}$($S_{2}$) goes to $D_{1}$($D_{2}$). Similarly, $A_{\text{exchange}}$ refers to the processes where particle from $S_{1}$($S_{2}$) goes to $D_{2}$($D_{1}$). The corresponding amplitudes are given as,
\begin{align}\label{eq:directandexchangeprocess}
    A_{\text{direct}} = \sum_{n,m} A_{n}^{(1\rightarrow1)}A_{m}^{(2\rightarrow2)} = A(1\rightarrow 1)A(2\rightarrow 2) \,. \\
    A_{\text{exchange}} = \sum_{n,m} A_{n+\frac{1}{2}}^{(1\rightarrow2)}A_{m+\frac{1}{2}}^{(2\rightarrow1)} = A(1\rightarrow 2)A(2\rightarrow 1) \,.
\end{align}
From here, we can simplify the probability, $P(11)_{\text{B/F}}$,
\begin{align}
  P(11)_{\text{B/F}}&=\mathcal{P}(1\rightarrow1){}^{2}+\mathcal{P}(1\rightarrow2){}^{2}\pm2\text{Re}[A_{\text{direct}}^{*}A_{\text{exchange}}]\,.\\
P(11)_{\text{Cl}} & =\mathcal{P}(1\rightarrow1)_{\text{Cl}}^{2}+\mathcal{P}(1\rightarrow2)_{\text{Cl}}^{2}\,.
\end{align}
The interference term in the probability, $P(11)_{\text{B/F}}$ is given as,
\begin{align}
    \text{Re}[A_{\text{direct}}^{*}A_{\text{exchange}}] & =-\frac{8r^{2}(1-r^{2})^{2}\sin^{2}\alpha}{(1+r^{2})^{2}}=-\frac{8(1-t^{2})t^{4}\sin^{2}\alpha}{(2-t^{2})^{2}}\,.
\end{align}
We shall take the limit of small $t$ and neglect terms of
order $t^4$ and higher. 
\begin{align}
\mathcal{P}(1\rightarrow1)_{\text{Cl}}&=1-\frac{t^{2}}{2} \;\text{and}\;\mathcal{P}(1\rightarrow2)_{\text{Cl}}=\frac{t^{2}}{2}\,.\\
\Delta_{1} & =c_{1}t^{2}\cos2\alpha \label{eq:delta1}\,.
\end{align}
For evaluating $P(11)_{\text{B/F}}$, we would again work in the limit
where $t\ll1$ and keep terms only of the order $(t^2)$,
\begin{align}
\mathcal{P}(1\rightarrow1)^{2} & =\big(1-\frac{t^{2}}{2}-c_{1}t^{2}\cos2\alpha\big)^{2}=1-t^{2}(1+c_{1}\cos2\alpha)\,.\\
\mathcal{P}(1\rightarrow2)^{2} & =t^{4}\big(\frac{1}{2}+c_{1}\cos2\alpha\big)^{2}=\mathcal{O}(t^{4})\,.\\
\text{Re}[A_{\text{direct}}^{*}A_{\text{exchange}}] & =-\frac{8(1-t^{2})t^{4}\sin^{2}\alpha}{(2-t^{2})^{2}}=-2t^{4}\sin^{2}\alpha=\mathcal{O}(t^{4})\,.\\ \label{eq:partialinterf}
\mathcal{P}(1\rightarrow1)_{\text{Cl}}^{2} & =1-t^{2}\;\text{and}\;\mathcal{P}(1\rightarrow2)_{\text{Cl}}^{2}=\frac{t^{4}}{4}=\mathcal{O}(t^{4})\,.
\end{align}
From here we get the apparent bunching of fermions, Fig.~\ref{fig:p11andg2}a,b,
\begin{equation}
P(11)_{\text{B/F}}=1-t^{2}(1+2c_{1}\cos2\alpha)\leq1-t^{2}=P(11)_{\text{Cl}} \ \ \text{for} \ \ \cos2\alpha \geq 0\,.
\end{equation}
In other words, if we define the benchmark of bunching as $\mathcal{B}_{1}=P(11)_{\text{Cl}}$, then we obtain,
\begin{align}\label{eq:clbenchmark}
    \mathcal{Q}_{\text{Cl}} = P(11)_{\text{B/F}} - P(11)_{\text{Cl}}\leq 0\,,
\end{align} for certain values of the tunneling amplitude and hence an apparent bunching of fermions, Fig.~\ref{fig:p11andg2}a,b. Now, we would like to pin down the processes which leads to an apparent boson-like behavior of fermions. The single particle amplitude have the following leading behavior in $t^2$,
\begin{align}
A_{0}^{(1\rightarrow1)} & \sim\mathcal{O}(1)\,,\quad A_{m}^{(1\rightarrow1)}\sim\mathcal{O}(t^{2})\;\text{for}\:m\neq0\,,\quad\text{and}\quad A_{n+\frac{1}{2}}^{(1\rightarrow2)}\sim\mathcal{O}(t^{2})\;\text{for all}\;n\,.
\end{align}
From here we can write down the leading behavior of all the terms in the amplitude for exchange process $A_{\text{exchange}}$ as,
\begin{align}
A_{n+\frac{1}{2}}^{(1\rightarrow2)}A_{m+\frac{1}{2}}^{(1\rightarrow2)} & \sim\mathcal{O}(t^{4})\in A_{\text{exchange}} \ \ \text{for all} \ \ n,m \,.
\end{align}
Similarly, in the amplitude, $A_{\text{direct}}$,
there are three types of terms,
\begin{align}
A_{0}^{(1\rightarrow1)}A_{0}^{(1\rightarrow1)} & \sim\mathcal{O}(1),\quad A_{0}^{(1\rightarrow1)}A_{n}^{(1\rightarrow1)}\sim\mathcal{O}(t^{2})\,,\quad\text{and}\quad A_{m}^{(1\rightarrow1)}A_{n}^{(1\rightarrow1)}\sim\mathcal{O}(t^{4}) \ \ \text{for all} \ \ n,m\neq0\,.
\end{align}
From here, we can say that when $t\ll1$, the contributing terms are of the type $A_{0}A_{0}$ and $A_{0}A_{n}$ and it leads to apparent bunching of fermions. Similarly, in the case of non-identical junction $S-$matrix, Eq.~\eqref{eq:nonidenticaljunctions}, if we consider interference only between the processes where winding number differs at most by $1$, we get,
\begin{align}
    \Delta_1 = -\frac{2(1-r_{1}^{2})(1-r_{2}^{2})r_{1}r_{2}\cos(\varphi-\theta_1-\theta_2)}{1-r_{1}^{2}r_{2}^{2}} \,.
\end{align}
By using the same set of arguments as for the identical junctions we obtain apparent bunching of fermions for $t_{1},t_{2}\ll 1$.

\section{Wave packet emitted from a dilutor quantum point contact } \label{sec:dilutor}
In this section, we calculate the wave packet shape in the case where it is emitted by a dilutor quantum point contact (QPC) in contact with the source region of the one-dimensional channel. 
In order to emit an electron wave packet towards the collider, we consider a source QPC at a chemical potential $\mu + eV$ and the collider leads at chemical potential $\mu=vk_{f}$, where $k_{f}$ is the Fermi-momentum. Thus, the width of the emitted wave packet in momentum space is determined by the voltage $V$.
The emitted wave packet is of the following form, 
\begin{align}\label{eq:squarewp}
    \phi(k) = \sqrt{\ell} \Theta ( k - k_{f}  ) \Theta \Big( k_{f}+\frac{eV}{v} - k\Big)  \,,  
\end{align} 
where the normalization factor $\ell=\frac{v}{eV}$ The wave packet is symmetric about the mean momentum $k_{0}=k_{f}+\frac{eV}{2v}$. As depicted in Fig.~1a, the scattered wave packet is a train of wavelets separated by $L$. With this wave packet, we would like to calculate the overlap of two wavelets separated in winding number by $|n-m|\neq 0$,
\begin{align}
f(|n-m|)&=\int_{-\infty}^{\infty}dx\phi(x+|n-m|L)^{*}\phi(x)\,,\\
& =\frac{\ell}{2\pi}\int_{k_{f}}^{k_{f}+\frac{1}{\ell}}dk_{1}dk_{2}\int_{-\infty}^{\infty}dxe^{ik_{1}x}e^{-ik_{2}x}e^{-ik_{2}|n-m|L}\,,\\
 & =\ell\int_{k_{f}}^{k_{f}+\frac{1}{\ell}}dk_{1}dk_{2}e^{-ik_{2}|n-m|L}\delta(k_{1}-k_{2})\,,\\
 & =\ell\int_{k_{f}}^{k_{f}+\frac{1}{\ell}}dke^{-ikl}\frac{(-i|n-m|L)}{(-i|n-m|L)}\,,\\
 & =\frac{i\ell}{|n-m|L}\Big[e^{-i(k_{f}+\frac{1}{\ell})|n-m|L}-e^{-ik_{f}|n-m|L}\Big]\,, \\
 & =\frac{ie^{-ik_{f}|n-m|L}}{ |n-m|}\frac{\ell}{L}\Big[e^{-i|n-m|\frac{L}{\ell}}-1\Big]  \,,\label{eq:interfsuppressionfunction}
\end{align} 
which characterizes the suppression of interference, as detailed in the main text. 

Although $\ell$ characterizes the decay of interference, the width of a single wave packet is strictly speaking not given by $\ell$. 
In real space, the wave packet is given as,
\begin{equation}
\tilde{\phi}(x) =\frac{e^{ik_{f}x}}{ix}\sqrt{\frac{\ell}{2\pi}}\Big[e^{i x/\ell }-1\Big]\,. \label{eq:wavepacketzerotemperature}
\end{equation}
However, due to the sharp Fermi surface in Eq.~(\ref{eq:squarewp}), the function $\tilde{\phi}(x)$ decays slowly, leading to a diverging variance $\langle x^2 \rangle$. 
Nevertheless, in practice we do get a wave packet that has a finite width in real space. One can understand this by considering two possible models: first, considering the QPC operating at finite temperature, $1/\beta$, and  second, where the QPC is  turned ON for a finite time $T$.

\subsection{Wave packet in case of non-zero temperature $\beta^{-1}$}\label{sec:finitetempwp}
Consider a QPC at a temperature $\beta^{-1} > 0$. As result of finite temperature, the wave packet emitted by the QPC will have smearing (at the end) unlike the rectangular wave packet, Eq.~\eqref{eq:squarewp}. The wave packet is given as follows,
\begin{align}
\phi(k) & =\frac{1}{\sqrt{\mathcal{N}}}\big[1-f_{F}(k-k_{f})\big]\big[f_{F}(k-k_{f}-\frac{eV}{v})\big]\,,
\end{align}
where $\mathcal{N}$ is the normalization factor which in the limit $\beta \to \infty$ is equal to $\frac{v}{eV}$ and $f_{F}$ refers to the Fermi function given as follows (we set $k_B = 1$ here), 
\begin{align}
    f_{F}(k) = \frac{1}{e^{\beta v k}+1}\,,
\end{align}
and $\lim_{\beta\to\infty}f_{F}(k) = \Theta(k)$. We would like to calculate the variance of the wave packet in the position space and therefore we begin by evaluating the wave packet in the position space,
\begin{align}
    \tilde{\phi}(x) &= \frac{1}{\sqrt{2\pi}} \int_{-\infty}^{\infty} dk e^{ikx} \phi(k) \,, \\ \label{eq:wprealspace}
    &=\frac{e^{ik_{f}x}}{\sqrt{2\pi\mathcal{N}}}\int_{-\infty}^{\infty}dke^{ikx}f_{F}(k-\frac{eV}{v}) + \frac{e^{ik_{f}x}}{\sqrt{2\pi\mathcal{N}}}\int_{-\infty}^{\infty}dke^{ikx}f_{F}(k-\frac{eV}{v})f_{F}(k)\,,\\
    &= W_{1} + W_{2}\,,
\end{align}
where $\mathcal{N}$ is the normalization factor for the wave packet. We first evaluate $W_{1}$ as follows,
\begin{align}
W_{1} & =\frac{e^{ik_{f}x}}{\sqrt{2\pi\mathcal{N}}}\int_{-\infty}^{\infty}dke^{ikx}\frac{1}{e^{\beta v k-\beta e V}+1}\,,\\
 & =\frac{e^{i\big(k_{f}+\frac{eV}{v}\big) x}}{\beta v\sqrt{2\pi\mathcal{N}}}\int_{-\infty}^{\infty}dq\frac{e^{i\frac{q}{\beta v}x}}{e^{q}+1}\,,
\end{align}
where we have made a substitution $q=\beta v k - \beta eV$. This integral can be performed by contour integration by noting that the Fermi function have poles on the imaginary axis. This yields,
\begin{align}
     W_{1} & =-2\pi i\frac{e^{i\big(k_{f}+\frac{eV}{v}\big)x} }{\beta v\sqrt{2\pi\mathcal{N}}}\frac{1}{2\sinh\frac{x\pi}{v\beta}}\Theta(x) - 2\pi i\frac{e^{i\big(k_{f}+\frac{eV}{v}\big)x}}{\beta v\sqrt{2\pi\mathcal{N}}}\frac{1}{2\sinh\frac{x\pi}{\beta v}}\Theta(-x)\,,\\
 & =-2\pi i\frac{e^{i\big(k_{f}+\frac{eV}{v}\big)x}}{\beta v\sqrt{2\pi\mathcal{N}}}\frac{1}{2\sinh\frac{x\pi}{\beta v}}\,.
\end{align}
The two terms are a result of closing the contour in the lower half plane and upper half plane. Similarly, 
\begin{align}
W_{2} & =\frac{e^{ik_{f}x}}{\sqrt{2\pi\mathcal{N}}}\int_{-\infty}^{\infty}dke^{ikx}\frac{1}{e^{\beta v k-\beta eV}+1}\frac{1}{e^{\beta v k}+1}=\frac{e^{ik_{f}x}}{\beta v\sqrt{2\pi\mathcal{N}}}\int_{-\infty}^{\infty}dqe^{i\frac{q}{\beta v}x}\frac{1}{e^{q-\beta eV }+1}\frac{1}{e^{q}+1}\,,\\
 & =\frac{\Theta(x)e^{ik_{f}x}}{\beta v \sqrt{2\pi\mathcal{N}}}2\pi i\sum_{n=0}^{\infty}\lim_{q\to eV\beta + i(2n+1)\pi}e^{i\frac{q}{\beta v}x}\frac{1}{e^{q}+1}\frac{q-\beta eV-i(2n+1)\pi}{e^{q-\beta eV}+1}\\
 & -\frac{\Theta(-x)e^{ik_{f}x}}{\beta v\sqrt{2\pi\mathcal{N}}}2\pi i\sum_{n=0}^{\infty}\lim_{q\to eV\beta - i(2n+1)\pi}e^{i\frac{q}{\beta v}x}\frac{1}{e^{q}+1}\frac{q+\beta eV +i(2n+1)\pi}{e^{q-\beta eV}+1}\\
 & \;+\frac{\Theta(x)e^{ik_{f}x}}{\beta v\sqrt{2\pi\mathcal{N}}}2\pi i\sum_{n=0}^{\infty}\lim_{q\to i(2n+1)\pi}e^{i\frac{q}{\beta v}x}\frac{1}{e^{q-\beta eV}+1}\frac{q-i(2n+1)\pi}{e^{q}+1}\\
& \;-\frac{\Theta(-x)e^{ik_{f}x}}{\beta v\sqrt{2\pi\mathcal{N}}}2\pi i\sum_{n=0}^{\infty}\lim_{q\to -i(2n+1)\pi}e^{i\frac{q}{\beta v}x}\frac{1}{e^{q-\beta eV}+1}\frac{q+i(2n+1)\pi}{e^{q}+1}\,.
\end{align}
We have four contributions, coming from contour integration with poles on the imaginary axis (lower and upper half plane). We obtain,
\begin{align}
 W_{2} 
 & =-\frac{2\pi i}{\beta v\sqrt{2\pi\mathcal{N}}}\frac{e^{i\big(\frac{eV}{v}+k_{f}\big)x}}{1-e^{\beta eV}}\frac{1}{2\sinh\frac{x\pi}{\beta v}}-\frac{2\pi i}{\beta v\sqrt{2\pi\mathcal{N}}}\frac{e^{ik_{f}x}}{1-e^{-\beta eV}}\frac{1}{2\sinh\frac{x\pi}{\beta v}}\,.
\end{align}
Compiling all the results we get the following,
\begin{align}
\tilde{\phi}(x) & = W_{1} + W_{2}= \frac{1}{\sqrt{2\pi\mathcal{N}}}\int_{-\infty}^{\infty}dke^{ikx}f_{F}\big(k-\frac{eV}{v}\big)-\int_{-\infty}^{\infty}dke^{ikx}f_{F}\big(k-\frac{eV}{v}\big)f(k) \,,\\
 & =-2\pi i\frac{e^{i\big(\frac{eV}{v}+k_{f}\big)x}}{\beta v\sqrt{2\pi\mathcal{N}}}\frac{1}{2\sinh\frac{x\pi}{\beta v}}+\frac{2\pi i}{\beta\sqrt{\mathcal{N}}}\frac{e^{i\big(\frac{eV}{v}+k_{f}\big)x}}{1-e^{eV\beta v}}\frac{1}{2\sinh\frac{x\pi}{\beta v}}+\frac{2\pi i}{\beta v\sqrt{\mathcal{N}}}\frac{e^{ik_{f}x}}{1-e^{-eV\beta}}\frac{1}{2\sinh\frac{x\pi}{\beta v}}\,,\\
& =\frac{\pi i}{\beta v\sqrt{2\pi\mathcal{N}}}\frac{e^{ik_{f}x}}{1-e^{-eV\beta}}\frac{1}{\sinh\frac{x\pi}{\beta v}}\Big[1-e^{i\frac{eV}{v}x}\Big]\,.
\end{align}
When $\beta eV \gg 1$, we recover Eq.~(\ref{eq:wavepacketzerotemperature}): 
\begin{align}
\tilde{\phi}(x) & =-\frac{\pi i e^{ik_{f}x}}{\beta v\sqrt{2\pi\mathcal{N}}}\frac{\beta v}{x\pi}\Big[e^{i\frac{eV}{v}x}-1\Big]=\frac{e^{ik_{f}x}}{ix}\sqrt{\frac{\ell}{2\pi}}\Big[e^{i\frac{eV}{v}x}-1\Big] \,.
\end{align}
Let us calculate the expectation value of $x^{2}$,
\begin{align}
\langle \hat{x}^{2}\rangle & =\int_{-\infty}^{\infty}dxx^{2}|\tilde{\phi}(x)|^{2}\,,\\
& =\frac{4\pi^{2}}{\beta^{2}v^{2}\mathcal{N}\bigg(1-e^{-eV\beta}\bigg)^{2}}\int_{-\infty}^{\infty}dxx^{2}\frac{\sin^{2}\frac{eV}{2v}x}{\sinh^{2}\frac{x\pi}{\beta v}}\,,\\
 & =\frac{4\pi^{2}}{\beta^{2}v^{2}\mathcal{N}\bigg(1-e^{-eV\beta}\bigg)^{2}}\bigg(\frac{2v}{eV}\bigg)^{3}\frac{\pi^{2}(\frac{2\pi}{eV\beta}+3(\frac{2\pi}{eV\beta}-\pi\coth\frac{eV\beta }{2})\sinh^{-2}\frac{eV\beta }{2})}{6(\frac{2\pi}{eV\beta})^{4}}\,.\label{eq:xsquaremathematica}
\end{align}
where we have used Wolfram Mathematica to obtain the Eq.~\eqref{eq:xsquaremathematica}. In the case when  $\frac{\beta v}{\ell}\gg1$ (and thus $\beta eV \gg 1$) we get the following simplifications,
\begin{align}\label{eq:varinwp}
\langle \hat{x}^{2}\rangle & \approx \frac{4\pi^{2}}{\beta^{2}v^{2}\mathcal{N}}\bigg(\frac{2v}{eV}\bigg)^{3}\frac{\pi^{2}}{6}\bigg(\frac{eV\beta}{2\pi}\bigg)^{3}=\frac{2\pi\beta v}{3\mathcal{N}}=\frac{2\pi\beta v^{2}}{3eV}=\frac{2\pi\beta v \ell}{3}\,.
\end{align}
with $\mathcal{N}\approx\frac{eV}{v}=\frac{1}{\ell}$ (for $\beta eV\gg1$) and note that we have $\langle x \rangle = 0$ because $|\tilde{\phi}(x)|^{2}$ is an even function of $x$. Therefore the wave packet width in position space is 
$\Delta x=\sqrt{\langle x^{2}\rangle-\langle x\rangle^{2}}\approx \sqrt{\frac{2\pi\beta v\ell}{3}}$.

\subsection{Wave packet in case of finite operating window $T$ of the QPC}
In a real experimental setup when the source is turned ON for a duration $T$, then the maximum spread of the wave packet is restricted by $vT$, where $v$ is the velocity of the wave packet. We consider the following wave packet,
\begin{align}
    \phi(x) = \left\{\begin{array}{cc}
        \frac{e^{ik_{f}x}}{ix}\sqrt{\frac{\ell}{2\pi\mathcal{N}}}\Big[e^{i x/\ell }-1\Big] & \text{for $|x|\leq\frac{vT}{2}$} \\
        0 & \text{for $|x|>\frac{vT}{2}$}
    \end{array} \right. 
\end{align}
where $\mathcal{N}$ is the normalization of the wave packet in the position space. If the duration for which the source is turned ON is sufficiently large, then in that case normalization is unchanged and we have $\mathcal{N}=1$. Using this wave packet, we are going to calculate the variance. This is evaluated as follows,
\begin{align}
    \langle x^{2} \rangle &= \int_{-\frac{vT}{2}}^{\frac{vT}{2}} dx x^{2} \frac{\ell}{2\pi\mathcal{N} x^{2}}\bigg[ 2-2 \cos\big( \frac{x}{\ell} \big) \bigg] \,,\\
    &=\frac{\ell}{\pi}\bigg( vT - 2\ell \sin\frac{vT}{2\ell} \bigg)\,.
\end{align}
In the limit where $\frac{vT}{2\ell}\gg 1 $, we have $\mathcal{N}\to 1$ because $|\phi(x)|^{2}$ falls as $x^{-2}$ and therefore the normalization of the wave packet doesn't change and we have $\mathcal{N}\to1$. This implies, 
\begin{align}
    \langle x^{2} \rangle \approx \sqrt{\frac{\ell vT}{\pi}}\,.
\end{align}
From here obtain,
\begin{align}
    \frac{\sqrt{\langle x^{2} \rangle}}{vT} = \sqrt{\frac{\ell}{\pi v T}} \ll 1\,.
\end{align}
hence, the spatial size of the wave packet is much smaller than $vT$  as one would expect.

\section{Benchmark for extracting the true mutual statistics of particles}\label{sec:quantumstatistics}
In this section, we are going to extract the quantum statistics of the colliding particles by defining the benchmark, $\mathcal{B}_{2}$. Outcome of the two-particle processes is affected by the underlying statistics of the particles and by choosing the correct observable we can extract the statistical information. Probabilities of different single particle events are independent of statistics because of the absence of any statistical interactions due to other particles. Statistics of the underlying particles affects all processes where we have at least two particles in the collider. However, due to the non-point-like geometry we find that there are infinite number of possible trajectories for any single particle event, Fig.~\ref{fig:amplitudes}a-b, and self interference of such trajectories leads to an apparent bunching of fermions as we discussed in the main text and Sec.~\ref{sec:toymodel}. Therefore, in order to extract the true information about the mutual statistics we define, $\mathcal{Q}_{\text{Irr}}$,
\begin{equation}\label{eq:irbenchmark}
\mathcal{Q}_{\text{Irr}}=P(11)_{s_{1}s_{2}}-\mathcal{P}(1\rightarrow1)_{s_{1}}\mathcal{P}(2\rightarrow2)_{s_{2}}-\mathcal{P}(1\rightarrow2)_{s_{1}}\mathcal{P}(2\rightarrow1)_{s_{2}} = P(11)_{s_{1}s_{2}}-\mathcal{B}_{2}\,,
\end{equation}
where $\mathcal{B}_{2}=\mathcal{P}(1\rightarrow1)_{s_{1}}\mathcal{P}(2\rightarrow2)_{s_{2}}+\mathcal{P}(1\rightarrow2)_{s_{1}}\mathcal{P}(2\rightarrow1)_{s_{2}}$ and $P(11)_{s_{1}s_{2}}$ refers to the probability when each of the detector receives one particle with both the sources $S_1$ and $S_2$ active (subscripts denote the active sources). The single particle probability $\mathcal{P}(1\rightarrow 1)_{s_1}$ and $\mathcal{P}(1\rightarrow 2)_{s_1}$ corresponds to the case where we have only single active source $S_1$. The subscript ``Irr" means irreducible as we subtract the single-particle contributions from $P(11)$. By defining $\mathcal{Q}_{\text{Irr}}$ in this way we extract the terms which are only sensitive to the processes which involves exchange of particles and therefore have the information of the quantum statistics. For classical particles we obtain,
\begin{align}
\mathcal{Q_{\text{Irr;Cl}}} & =P(11)_{s_{1}s_{2}}-\mathcal{P}(1\rightarrow1)_{s_{1}}\mathcal{P}(2\rightarrow2)_{s_{2}}-\mathcal{P}(1\rightarrow2)_{s_{1}}\mathcal{P}(2\rightarrow1)_{s_{2}}\,,\\
 & =\mathcal{P}(1\rightarrow1)_{s_{1}}\mathcal{P}(2\rightarrow2)_{s_{2}}+\mathcal{P}(1\rightarrow2)_{s_{1}}\mathcal{P}(2\rightarrow1)_{s_{2}}\\
 & \quad-\mathcal{P}(1\rightarrow1)_{s_{1}}\mathcal{P}(2\rightarrow2)_{s_{2}}-\mathcal{P}(1\rightarrow2)_{s_{1}}\mathcal{P}(2\rightarrow1)_{s_{2}}\,,\\
 &=0\,,
\end{align}
as expected because classical particles doesn't have any quantum statistics. With bosons and fermions, we have,
\begin{align}
P(11)_{s_{1}s_{2}} & =|A_{\text{direct}}|^{2}+|A_{\text{exchange}}|^{2}\pm2\text{Re}\big[A_{\text{direct}}^{*}A_{\text{exchange}}\big]\,,
\end{align}
where the $+/-$ sign corresponds to bosons/fermions. The contribution of the direct processes and the exchange processes can be simplified as follows using Eq.~\eqref{eq:directandexchangeprocess},
\begin{align}
A_{\text{direct}} & =A(1\rightarrow1)A(2\rightarrow2) \ \ \text{and} \ \ A_{\text{exchange}}=A(1\rightarrow2)A(2\rightarrow1)\,,\\
|A_{\text{direct}}|^{2} & =\mathcal{P}(1\rightarrow1)_{s_{1}}\mathcal{P}(2\rightarrow2)_{s_{2}} \ \ \text{and} \ \ |A_{\text{exchange}}|^{2}=\mathcal{P}(1\rightarrow2)_{s_{1}}\mathcal{P}(2\rightarrow1)_{s_{2}}\,.
\end{align}
Therefore, we find that the benchmark $\mathcal{B}_{2}=\mathcal{P}(1\rightarrow1)_{s_{1}}\mathcal{P}(2\rightarrow2)_{s_{2}}+\mathcal{P}(1\rightarrow2)_{s_{1}}\mathcal{P}(2\rightarrow1)_{s_{2}}$ removes all the single-particle self-interference terms and hence we have,
\begin{align}
\mathcal{Q}_{\text{Irr;B/F}}=\pm2\text{Re}\big[A_{\text{direct}}^{*}A_{\text{exchange}}\big]\,.
\end{align}
The function $\mathcal{Q}_{\text{Irr}}$ gives us the information about the statistics of the particles. It comes with specific signature for bosons and fermions and hence distinguishes the statistical nature of bosons and fermions. The earlier defined $\mathcal{Q}_{\text{Cl}}$ didn't remove the self interference contributions to the probability $P(11)_{\text{F/B}}$ and hence was giving false information about the statistics of the colliding particles. We emphasize that $\mathcal{Q}_{\text{Cl}}$, Eq.~\eqref{eq:clbenchmark} is \textit{not} the correct benchmark to extract the mutual statistics of particles. We summarize the results for the probabilities for classical particles, classical waves, quantum particles and quantum waves with both, single active source and two active sources in the table, Tab.~\ref{tab:summary}. 
\begin{table} \label{tab:summary}
    \centering
    \renewcommand\arraystretch{1.5}
    \resizebox{\textwidth}{!}{%
    \begin{tabular}{|c|c|c|c|c|}
        \hline
         & \multicolumn{2}{c|}{Distinguishable particle(s)} & \multicolumn{2}{c|}{Indistinguishable particle(s)}  \\ \hline
         & Classical particle(s) & Classical wave(s) & Quantum particle(s) & Quantum wave(s) \\ \hline 
         \makecell{One active \\ source $(S_1)$} & $\begin{array}{c}
         \mathcal{P}(1 \rightarrow 1)_{S_1} = \sum_{n=0}^{\infty} |A_{n}^{(1 \rightarrow 1)}|^2 \\
         \mathcal{P}(1 \rightarrow 2)_{S_1} = \sum_{n=0}^{\infty} |A_{n+\frac{1}{2}}^{(1 \rightarrow 2)}|^2
         \end{array}$ &  $\begin{array}{c}
         \mathcal{P}(1 \rightarrow 1)_{S_1} = \big|\sum_{n=0}^{\infty} A_{n}^{(1 \rightarrow 1)}\big|^2 \\
         \mathcal{P}(1 \rightarrow 2)_{S_1} = \big|\sum_{n=0}^{\infty} A_{n+\frac{1}{2}}^{(1 \rightarrow 2)}\big|^2
         \end{array}$ & $\begin{array}{c}
         \mathcal{P}(1 \rightarrow 1)_{S_1} = \sum_{n=0}^{\infty} |A_{n}^{(1 \rightarrow 1)}|^2 \\
         \mathcal{P}(1 \rightarrow 2)_{S_1} = \sum_{n=0}^{\infty} |A_{n+\frac{1}{2}}^{(1 \rightarrow 2)}|^2
         \end{array}$  & $\begin{array}{c}
         \mathcal{P}(1 \rightarrow 1)_{S_1} = \big|\sum_{n=0}^{\infty} A_{n}^{(1 \rightarrow 1)}\big|^2 \\
         \mathcal{P}(1 \rightarrow 2)_{S_1} = \big|\sum_{n=0}^{\infty} A_{n+\frac{1}{2}}^{(1 \rightarrow 2)}\big|^2
         \end{array}$ \\ \hline
         \makecell{Two active \\ sources ($S_1$ and $S_2$)}  & $\begin{array}{rcl}
P(11) & = & \mathcal{P}(1\to 1)_{S_1}\mathcal{P}(2\to 2)_{S_2} \\
      &   & + \ \mathcal{P}(1\to 2)_{S_1}\mathcal{P}(2\to 1)_{S_2}
\end{array}$ & $\begin{array}{rcl}
P(11) & = & \mathcal{P}(1\to 1)_{S_1}\mathcal{P}(2\to 2)_{S_2} \\
      &   & + \ \mathcal{P}(1\to 2)_{S_1}\mathcal{P}(2\to 1)_{S_2}
\end{array}$ & $\begin{array}{rcl}
P(11) & = & \sum_{n,m=0}^{\infty} \big|A_{n}^{(1\to1)}A_{m}^{(2\to2)} \\
      &   & \pm A_{n+\frac{1}{2}}^{(1\to2)}A_{m+\frac{1}{2}}^{(2\to1)}\big|^{2} \\
      & \approx & \mathcal{P}(1\to 1)_{S_1}\mathcal{P}(2\to 2)_{S_2} \\
      &   & + \ \mathcal{P}(1\to 2)_{S_1}\mathcal{P}(2\to 1)_{S_2}
\end{array}$ & $\begin{array}{rcl}
P(11)_{\text{B/F}} & = &  \mathcal{P}(1\to 1)_{S_1}\mathcal{P}(2\to 2)_{S_2} \\
      &   & +\mathcal{P}(1\to 2)_{S_1}\mathcal{P}(2 \to 1)_{S_2} \\
      &   & \pm 2\text{Re}\Big[ \sum_{n,m,i,j=0}^{\infty} A_{n}^{(1\to 1)}A_{m}^{(2\to 2)} \\
      &   & 
     \times A^{(1\to2)*}_{i+\frac{1}{2}}A^{(2\to1)*}_{j+\frac{1}{2}} \Big]  
      \end{array}$\\
      \hline
      \makecell{Comment on the two \\ benchmarks of bunching} & $\begin{array}{rcl}
          \mathcal{Q}_{\text{Cl}} & = & 0 \ \ \text{for all $r$}  \\
           \mathcal{Q}_{\text{Irr}} & = & 0 \ \ \text{for all $r$}
      \end{array}$ & $\begin{array}{rcl}
          \mathcal{Q}_{\text{Cl}} & \leq & 0 \ \ \text{for some $r$} \\
           \mathcal{Q}_{\text{Irr}} & = & 0 \ \ \text{for all $r$}
      \end{array}$ & $\begin{array}{rcl}
          \mathcal{Q}_{\text{Cl}} & = & 0 \ \ \text{for all $r$} \\
           \mathcal{Q}_{\text{Irr}} & = & 0 \ \ \text{for all $r$}
      \end{array}$ & $\begin{array}{rcl}
          \mathcal{Q}_{\text{Cl;B/F}} & \leq & 0 \ \ \text{for some $r$}  \\
           \mathcal{Q}_{\text{Irr;B}} & \leq & 0 \ \ \text{for all $r$} \\
           \mathcal{Q}_{\text{Irr;F}} & \geq & 0 \ \ \text{for all $r$}
      \end{array}$ \\
      \hline
    \end{tabular}
    }
    \caption{
    Probabilities of different events in the extended collider with classical particles, classical waves, quantum particles and quantum waves with one or two active sources. Differences between them arises from the degree of interference they have, with classical particles there is no interference at all and with quantum waves, we have full interference (everything interferes with everything else). In the table, by quantum particles, we mean a wave packet that is localized to a region smaller than the size of the quantum anti-dot ($L$), 
    which leads to less interference and therefore the probability is similar to that of classical particles. In the limit of very small width ($\ell \ll L$) we recover the classical particles probability. For extracting the statistical differences we have found that there are two benchmarks. In one, we calculate the probability, $P(11)$ and compare it with the corresponding quantity for classical particles, which doesn't remove the single particle self-interference terms and the other involves defining an irreducible part of $P(11)$ which removes the contributions from single particle interference. If we choose the first benchmark of bunching, the corresponding probability of classical particles, $P(11)_{\text{Cl}}$ and define, $\mathcal{Q}_{\text{Cl}}=P(11)-P(11)_{\text{Cl}}$, Eq.~\eqref{eq:clbenchmark}. Then, we get an apparent bunching of fermions, $\mathcal{Q}_{\text{Cl;F}}\leq 0$ for certain values of $r$, Fig.~2a,b. With the second benchmark, we define the irreducible correlator, $\mathcal{Q}_{\text{Irr}}
    =P(11)_{S_{1}S_{2}}-\mathcal{P}(1\rightarrow1)_{S_{1}}\mathcal{P}(2\rightarrow2)_{S_{2}}-\mathcal{P}(1\rightarrow2)_{S_{1}}\mathcal{P}(2\rightarrow1)_{S_{2}}$, Eq.~\eqref{eq:irbenchmark}, one can differentiate bosons ($\mathcal{Q}_{\text{Irr}}\leq 0$) from fermions ($\mathcal{Q}_{\text{Irr}}\geq 0$) for all values of $r$.} 
\end{table}
Quantum particles mentioned here refers to the narrow wave packets of indistinguishable particles. Width of the wave packet controls the degree of the interference in the system. With extremely narrow wave packet ($\sqrt{\beta v \ell}/L \ll 1$), Eq.~\eqref{eq:varinwp}, there is no interference at all and we recover the classical particle results. In the proposed model of the extended collider, for quantum particles (narrow wave packets), we have,
\begin{align}
    P(11)_{\text{B/F}} =& \sum_{n,m=0}^{\infty} |A_{n}^{(1\to1)}A_{m}^{(2\to2)} \pm A_{n+\frac{1}{2}}^{(1\to1)}A_{m+\frac{1}{2}}^{(2\to2)}|^{2}\,, \\
    =& \sum_{n=0}^{\infty}|A_{n}^{(1\to1)}|^{2}\sum_{m=0}^{\infty}|A_{m}^{(2\to2)}|^{2} + \sum_{n=0}^{\infty}|A_{n+\frac{1}{2}}^{(1\to2)}|^{2}\sum_{m=0}^{\infty}|A_{m+\frac{1}{2}}^{(2\to1)}|^{2} \pm 2\text{Re}\Big[ A_{n}^{(1\to1)}A_{n+\frac{1}{2}}^{(1\to2)}A_{m}^{(2\to2)}A_{m+\frac{1}{2}}^{(2\to1)} \Big]\,, \\
    \approx& \sum_{n=0}^{\infty}|A_{n}^{(1\to1)}|^{2}\sum_{m=0}^{\infty}|A_{m}^{(2\to2)}|^{2} + \sum_{n=0}^{\infty}|A_{n+\frac{1}{2}}^{(1\to2)}|^{2}\sum_{m=0}^{\infty}|A_{m+\frac{1}{2}}^{(2\to1)}|^{2}\,.
\end{align}With narrow wave packets ($\sqrt{\beta v \ell}/L \ll 1$), Eq.~\eqref{eq:varinwp}, we are going to have,
\begin{align}
    A_{n}^{(1\to1)}A_{n+\frac{1}{2}}^{(1\to2)} \rightarrow 0\,,
\end{align} 
which can be understood from the fact that a wave packet from $S_1$ going to $D_2$ will have to travel an extra distance of half the loop size $(L/2)$, and with narrow wave packet, its interference with wave packet from $S_{1}$ to $D_{1}$ is very small and can be neglected. Therefore, the differences that we see in the probabilities is a result of difference in the level of interference coming from the size of the wave packet. The level of interference can be varied by changing the spatial width of the wave packet.

\section{Estimation of spatial width of the scattered wave packet and the condition on the emission rate at the sources}\label{sec:widthscatteredwave}
An incoming wave packet after scattering, becomes an infinite series of wavelets in the transmission/reflection lead, Fig.~1a. The center of each of the wavelet is separated by $L$. Since, our calculations requires no interference between wave packets from the same source, we would like to calculate the spatial width of the transmitted/reflected wave packets and then determine the correct emission rate of the sources. In the following we present an approximate calculation of the width of the scattered wave packet. 
By using Eq.~(2) and Eq.~(3), we have, 
\begin{equation}\label{eq:transmittedwavelet}
\tilde{\phi}_{d_{1}}(x)=\rho_{1}\tilde{\phi}_{s_{1}}(x)+\rho_{2}\tau_{1}\tau_{2}\tilde{\phi}_{s_{1}}(x+L)+\rho_{2}^{3}\tau_{1}\tau_{2}\tilde{\phi}_{s_{1}}(x+2L)+\cdots \,.
\end{equation}
For calculating the variance of the transmitted wave packet, we first need to evaluate the expectation value of the position operators, $\hat{x}$ and $\hat{x}^{2}$,
\begin{align}
\langle\hat{x}\rangle & =\int_{-\infty}^{\infty}dxx|\tilde{\phi}_{d_{1}}(x)|^{2}  \ \ \text{and} \ \ \langle\hat{x}^{2}\rangle =\int_{-\infty}^{\infty}dxx^{2}|\phi_{d_{1}}(x)|^{2} \,.
\end{align}
Since, the transmitted wave packet have infinite number of terms, Eq.~\eqref{eq:transmittedwavelet}, calculation of the expectations value becomes non-trivial and therefore we do the calculations in the classical limit (by 'classical', we mean in this paper the complete lack of interference, still allowing for tunneling processes) and drop all the cross terms (interference terms). By only considering the classical contribution to the expectation values, we obtain for the transmitted wave packet,
\begin{align}
    \langle \hat{x} \rangle_{\text{trans}} &= |\rho_{1}|^{2}\int_{-\infty}^{\infty}dx|\tilde{\phi}_{s_{1}}(x)|^{2}x+|\rho_{2}\tau_{1}\tau_{2}|^{2}\int_{-\infty}^{\infty}dx|\tilde{\phi}_{s_{1}}(x+L)|^{2}x+\cdots \,,\\
    \langle \hat{x}^{2} \rangle_{\text{trans}} &= |\rho_{1}|^{2}\int_{-\infty}^{\infty}dx|\tilde{\phi}_{s_{1}}(x)|^{2}x^{2}+|\rho_{2}\tau_{1}\tau_{2}|^{2}\int_{-\infty}^{\infty}dx|\tilde{\phi}_{s_{1}}(x+L)|^{2}x^{2}+\cdots \,,
\end{align}
where the subscript ``trans" refers to the transmitted wave packet. Each of the term corresponds to diagram in the Fig.~\ref{fig:amplitudes}a and $\tilde{\phi}_{s_{1}}(x)$ is given by Eq.~\eqref{eq:wprealspace}. Note that the quantity $|\phi(x)|^{2}$ is symmetric around $x=0$ (even function) and therefore we have,
\begin{align}
    \int_{-\infty}^{\infty}dx|\tilde{\phi}_{s_{1}}(x+L)|^{2}x &=\int_{-\infty}^{\infty}dx|\tilde{\phi}_{s_{1}}(x)|^{2}(x-L)\,,\\
    &=-L\int_{-\infty}^{\infty}dx|\tilde{\phi}_{s_{1}}(x)|^{2}=-L\,.
\end{align}
Similarly, we have,
\begin{align}
    \int_{-\infty}^{\infty}dx|\tilde{\phi}_{s_{1}}(x+L)|^{2}x^{2} &=\int_{-\infty}^{\infty}dx|\tilde{\phi}_{s_{1}}(x)|^{2}(x-L)^{2}\,,\\
    &=L^{2}\int_{-\infty}^{\infty}dx|\tilde{\phi}_{s_{1}}(x)|^{2}+\int_{-\infty}^{\infty}dx|\tilde{\phi}_{s_{1}}(x)|^{2}x^{2}\approx L^{2} + \frac{2\pi\beta v \ell}{3}\,,
\end{align}
where we have used the fact that $\int_{-\infty}^{\infty}dx|\tilde{\phi}_{s_{1}}(x)|^{2}x^{2}=\frac{2\pi\beta v \ell}{3}$, Eq.~\eqref{eq:varinwp}, which is valid in the limit $\frac{\beta v}{\ell} \gg 1$. From here we obtain the following,
\begin{align}
    \langle \hat{x} \rangle_{\text{trans}} = -\frac{LR}{1+R^2}  \ \ \text{and} \ \ \langle \hat{x}^{2} \rangle_{\text{trans}} = \frac{2\pi\beta v \ell}{3}\mathcal{P}(1\to1)_{\text{Cl}} + \frac{L^{2}R(1+R^{2})}{(1-R)(1+R)^{3}} \,,
\end{align}
where we have $R=|\rho_{1}|^{2}=|\rho_{2}|^{2}$. A similar set of calculations can be done for the reflected wave packet. Below, we present the spatial width of the wave packet in the limit $\sqrt{\beta v \ell}/L \ll 1$,
\begin{align}
    L_{\text{trans}} &= \sqrt{\langle\hat{x}^{2}\rangle_{\text{trans}}-\langle\hat{x}\rangle_{\text{trans}}^{2}} \approx 
    \sqrt{\frac{L^2 R \big( R^3 + 2R^2 +1\big)}{(1-R)(1+R)^{4}}} = \sqrt{\frac{\tau_{0} v L R \big( R^3 + 2R^2 +1\big)}{(1+R)^{3}}}\,,\label{eq:transwidth}\\
 L_{\text{refl}} &= \sqrt{\langle\hat{x}^{2}\rangle_{\text{refl}}-\langle\hat{x}\rangle^{2}_{\text{refl}}}\approx
    \sqrt{\frac{L^2 R \big(1 + R^4 + 4R^2 +2R\big)}{2(1-R)(1+R)^{4}}} = \sqrt{\frac{\tau_{0}vL R \big(1 + R^4 + 4R^2 +2R\big)}{2(1+R)^{3}}} \,,  \label{eq:reflwidth}
\end{align}
where we have $\tau_{0}$ is the lifetime of the particle in the QAD. Lifetime is defined as the average time that the particle spends inside the QAD before it escapes. Lets say we created a particle inside the QAD, probability of performing one winding before escaping is $R^{2}$, probability of performing two winding before escaping is $R^{4}$ and so on. The corresponding time spent on the QAD is $L/v$ for performing one winding and $2L/v$ for performing two winding and so on. This implies the lifetime of the particle is given as
\begin{align}\label{eq:lifetime}
    \tau_{0} =& \frac{1}{N}\Big[\frac{R^{2}L}{v} + \frac{2R^{4}L}{v} + \frac{3R^{6}L}{v} + \cdots\Big]= \frac{L}{v(1-R^{2})}\,,
\end{align}
where $N$ is the normalization factor and is equal to the expectation value of $1$ with respect to the probability of escaping after performing $n$ number of winding, i.e. $N=\sum_{n=1}^{\infty}R^{2n}$. We again emphasize that the calculations presented in this section are in the limit when the wave packets are well localized, $\frac{\sqrt{\beta v \ell}}{L} \ll 1$. 

Since we have assumed that the emission rate at the sources is $\lambda$, this implies that the average time difference between the emission of consecutive particles is $\frac{1}{\lambda}$. Note that our analysis presented in the manuscript assumes no more than two particles simultaneously in the collider and this can be achieved when we have the average distance between the particles is much larger than the width of the scattered wave packet ($\sim L$ see Eq.~\eqref{eq:transwidth} and Eq.~\eqref{eq:reflwidth}), i.e. $\frac{v}{\lambda}\gg L$. This is the condition on $\lambda$ as mentioned in the main text. 

\subsection{Comment on the case with many particles emitted from the source}
Here, we would like to see the case where we have multiple particles simultaneously in the collider emitted from the source(s). We begin by estimating the probability of having two particles within the time interval $\epsilon \sim \frac{L}{v}$ given the source, say $S_{1}$ emitted $n$ number of particles. So, the probability of having two particles within the time interval $\epsilon$ with emission of $n$ number particle, $P_{n}^{(2)}$,
\begin{align}
    P_{n}^{(2)} & =\binom{n}{2}\frac{\epsilon(2T-\epsilon)}{T^{2}}\,,\\
 & =\frac{n(n-1)}{2}\frac{(2T\epsilon-\epsilon^{2})}{T^{2}}\,.
\end{align}
where $\epsilon(2T-\epsilon)$ corresponds to the area of the strip in the phase space where the particles are within time interval $\epsilon$. Let us assume that $\epsilon$ is small compared to $T$ and therefore we can neglect
$\epsilon^{2}$ terms,
\begin{equation}
P_{n}^{(2)}=\frac{\epsilon}{T}n(n-1).
\end{equation}
In the case where we have lots of particles emitted from the source in a given time interval $T$, we can approximate the above calculated probability as follows,
\begin{align}
    P_{n}^{(2)}\approx\frac{\epsilon}{T}\lambda^{2}T^{2}=\epsilon \lambda^{2} T,
\end{align}
where we have used the fact that on average we have $n=\lambda T$ number of particles from the source(s). Generalizing this formula to find the probability of having three particles within the time interval $\epsilon$ is given as,
\begin{align}
    P_{n}^{(3)} \approx \frac{\epsilon^{2}Tn^{3}}{T^{3}}=\epsilon^{2}\lambda^{3}T.
\end{align}
From the above calculations, we can see the probability of having more and more particles within a time interval is suppressed by $\epsilon\lambda$. So, in our case we choose a low enough emission rate ($\lambda \ll \frac{v}{L}$) and therefore the undesired multi-particle interference effects are small and can be ignored.

\section{Relation between the current-current correlations and probabilities}\label{sec:crosscurrentcorrelator}
In this section we are going to look at the  antibunching probability $P(11)_{\text{F/B}}$ of receiving one particle at each detector $D_{1}$ and $D_{2}$ with particles coming in from both the sources. We establish the relationship between the two-particle probability $P(11)_{\text{F/B}}$ and the cross-current correlation function measured at the detectors, $D_{1}$ and $D_{2}$. This is an experimentally accessible quantity. In our system, current at the detector is proportional to the number
density at the detector and is given as $I_{d_{i}}(t_{i})=evn_{d_{i}}(t_{i})$
for $i=1,2$ and $e$ is the charge and $n_{d_i}(x,t)=\phi_{d_i}^{\dagger}(x,t)\phi_{d_i}(x,t)$ is the number density in terms of the field operator $\phi_{d_i}^{\dagger}(x,t)$. Let us begin by considering the case where we have just a single active
source and we try to compute the single particle probability $\mathcal{P}(1\to1)$.
Let us assume that the source $S_{1}$ is turned ON for the time interval $[-T/2,T/2]$ and in the end of the calculations, we set $T\to \infty$. Lets say that the detector, $D_{1}$ is at the coordinate $x_{1}$
and the measurement time is $t_{1}$. Now the incoming state is a mixed
state from the source $S_{1}$ and is given as,
\begin{equation}
\rho_{s_{1}}=\sum_{n_{1}=0}^{\infty}P_{\lambda}(n_{1})\Big[\prod_{i_{1}=1}^{n_{1}}\tilde{\phi}_{s_{1}}^{\dagger}(x_{1}^{(0)},t_{i_{1}}^{(0)})|\Omega\rangle\langle\Omega|\tilde{\phi}_{s_{1}}(x_{1}^{(0)},t_{i_{1}}^{(0)})\Big]\,,
\end{equation}
where we have assumed that the emitted particles follow a Poisson
distribution with emission rate $\lambda$. The probability
of emitting $n$ number of particles in the time interval $T$ from the source is given as,
\begin{equation}
P_{\lambda}(n)=\frac{e^{-\lambda T}(\lambda T)^{n}}{n!} \,.
\end{equation}
For concreteness let us consider a particular case where the incoming state
have $n_{1}$ number of particles from the source $S_{1}$ and using this, we will calculate
the number density at the detector $D_{1}$. The state with $n_{1}$
number of particles is given as,
\begin{equation}
|\psi_{n_{1}}\rangle_{s_{1}}=\prod_{i_{1}=1}^{n_{1}}\tilde{\phi}_{s_{1}}^{\dagger}(x_{1}^{(0)},t_{i_{1}}^{(0)})|\Omega\rangle \,.
\end{equation}
Number density evaluated at the detector $D_{1}$ at time $t_{1}$ is as follows,
\begin{equation}
\langle\psi_{n_{1}}|n_{d_{1}}(t_{1})|\psi_{n_{1}}\rangle_{s_{1}}=\Big|\sum_{i_{1}=1}^{n_{1}}\mathcal{I}_{1}(x_{1},t_{1};x_{1}^{(0)},t_{i_{1}}^{(0)})\Big|^{2} \,,
\end{equation}
where the subscript $s_{1}$ outside the angular bracket emphasizes that we work with a particular state, $|\psi_{n_{1}}\rangle_{s_{1}}$. For a dilute stochastic beam of particles we are going to have emission
times well separated (mean separation being $1/\lambda$) and therefore we neglect the interference terms between the particles that are emitted at different times. By integrating over the emission times, we obtain,
\begin{align}
v\int_{-T/2}^{T/2}\prod_{i_{1}=1}^{n_{1}}dt_{i_{1}}\langle\psi_{n_{1}}|n_{d_{1}}(t_{1})|\psi_{n_{1}}\rangle_{s_{1}} & =T^{n_{1}-1}n_{1}\mathcal{P}(1\to1)\,, \\ 
\frac{v}{T^{n_{1}}}\int_{-T/2}^{T/2}\prod_{i_{1}=1}^{n_{1}}dt_{i_{1}}\langle\psi_{n_{1}}|n_{d_{1}}(t_{1})|\psi_{n_{1}}\rangle_{s_{1}} & =\frac{n_{1}}{T}\mathcal{P}(1\to1) \,.
\end{align}
Averaging over number of particles from the source $S_{1}$ we have,
\begin{align}
\sum_{n_{1}=0}^{\infty}P_{\lambda}(n_{1})\frac{v}{T^{n_{1}}}\int_{-T/2}^{T/2}\prod_{i_{1}=1}^{n_{1}}dt_{i_{1}}\langle\psi_{n_{1}}|n_{d_{1}}(t_{1})|\psi_{n_{1}}\rangle_{s_{1}} & =\sum_{n_{1}=0}^{\infty}P_{\lambda}(n_{1})\frac{n_{1}}{T}\mathcal{P}(1\to1)\,, \\ 
 & =\lambda \mathcal{P}(1\to1) \,.
\end{align}
This implies that the average current at the drain with a single active
source, $S_{1}$, is given as,
\begin{equation}\label{eq:singlesourceprob1}
\langle I_{d_{1}}(t_{1})\rangle_{s_{1}}=e\lambda \mathcal{P}(1\to1)=\langle I_{d_{1}}\rangle_{s_{1}}\,, \quad(\text{a constant current in the drain $D_{1}$})
\end{equation}
and similarly in the case of active $S_{2}$, we have,
\begin{equation}\label{eq:singlesourceprob2}
\langle I_{d_{2}}(t_{2})\rangle_{s_{2}}=e\lambda \mathcal{P}(1\to2)=\langle I_{d_{2}}\rangle_{s_{2}}\,, \quad(\text{a constant current in the drain $D_{2}$})
\end{equation}
Now, let us generalize the above analysis with stochastic beam of
particles from the sources $S_{1}$ and $S_{2}$, both being simultaneously active. The
density matrix in this case is given as follows,
\begin{equation}\label{eq:stochasticbeams1s2}
\rho_{s_{1}s_{2}}=\sum_{n_{1},n_{2}=0}^{\infty}P_{\lambda}(n_{1})P_{\lambda}(n_{2})\Big[\prod_{i_{1}=1}^{n_{1}}\prod_{i_{2}=1}^{n_{2}}\tilde{\phi}_{s_{2}}^{\dagger}(x_{2}^{(0)},t_{i_{2}}^{(0)})\tilde{\phi}_{s_{1}}^{\dagger}(x_{1}^{(0)},t_{i_{1}}^{(0)})|\Omega\rangle\langle\Omega|\tilde{\phi}_{s_{1}}(x_{1}^{(0)},t_{i_{1}}^{(0)})\tilde{\phi}_{s_{2}}^{\dagger}(x_{2}^{(0)},t_{i_{2}}^{(0)})\Big] \,,
\end{equation}
where the subscript denotes the active sources. Let us focus on the
particular case where we have $n_{1}$ particles from the source $S_{1}$
and $n_{2}$ number of particles from the source $S_{2}$. The corresponding
quantum state is given as,
\begin{equation}
|\psi_{n_{1}n_{2}}\rangle_{s_{1}s_{2}}=\prod_{i_{1}=1}^{n_{1}}\prod_{i_{2}=1}^{n_{2}}\tilde{\phi}_{s_{2}}^{\dagger}(x_{2}^{(0)},t_{i_{2}}^{(0)})\tilde{\phi}_{s_{1}}^{\dagger}(x_{1}^{(0)},t_{i_{1}}^{(0)})|\Omega\rangle \,.
\end{equation}
Let us assume that the detector $D_{1}$ is at $x_{1}$ and $D_{2}$ is at $x_{2}$ and the measurement times are $t_{1}$ and $t_{2}$ correspondingly. Next, we calculate the density-density correlations at the detector $D_{1}$ and $D_{2}$ averaged over the emission times $\{t_{i_{1}}^{(0)},t_{i_{2}}^{(0)}\}$. We also assume that the time interval over which we average ($T$) is large, this
implies,
\begin{align}\label{eq:mixedstateprob}
\frac{v^{2}}{T^{n_{1}}T^{n_{2}}}\int_{-T/2}^{T/2}\prod_{i_{1}=1}^{n_{1}}dt_{i_{1}}\prod_{i_{2}=1}^{n_{2}}dt_{i_{2}}\langle n_{d_{1}}(t_{1})n_{d_{2}}(t_{2})\rangle_{n_{1}n_{2}} & =\frac{v^{2}}{T^{n_{1}}T^{n_{2}}}\int_{-T/2}^{T/2}\prod_{i_{1}=1}^{n_{1}}dt_{i_{1}}\prod_{i_{2}=1}^{n_{2}}dt_{i_{2}}\langle\psi_{n_{1}n_{2}}|n_{d_{1}}(t_{1})n_{d_{1}}(t_{2})|\psi_{n_{1}n_{2}}\rangle\,,\\
 & =\frac{n_{1}n_{2}}{T^{2}}P(11)_{\text{CW}}\mp\frac{2n_{1}n_{2}}{T^{2}}\mathcal{F}(x_{1},t_{1};x_{2},t_{2})\\
 & \quad+\frac{2}{T^{2}}\binom{n_{1}}{2}P(20)_{\text{CW}}\pm\frac{2}{T^{2}}\binom{n_{1}}{2}\mathcal{F}(x_{1},t_{1};x_{2},t_{2})\\
 & \quad+\frac{2}{T^{2}}\binom{n_{2}}{2}P(02)_{\text{CW}}\pm\frac{2}{T^{2}}\binom{n_{2}}{2}\mathcal{F}(x_{1},t_{1};x_{2},t_{2})\,,
\end{align}
where the subscript outside the angular brackets on LHS emphasizes that we are working with a particular state $|\psi_{n_{1}n_{2}}\rangle_{s_{1}s_{2}}$ and $-/+$ sign is for fermions/bosons with the function $\mathcal{F}$ given as,
\begin{align}
\mathcal{F}(x_{1},t_{1};x_{2},t_{2}) & =-\Big|\int_{-\infty}^{\infty} dk_{1}e^{ik_{1}(x_{1}+x_{2}+vt_{2}-vt_{1})}\mathcal{T}_{k_{1}}\mathcal{R}_{k_{1}}^{*}|\phi(k_{1})|^{2}\Big|^{2}\,.
\end{align}
In obtaining the Eq.~\eqref{eq:mixedstateprob}, we have assumed that emission times of the particles are well separated, which is a result of small enough emission rate, $\lambda$ (mean separation being $\frac{1}{\lambda}$) and hence at any instant in time there are no more than two particles in the collider. Now, from Eq.~\eqref{eq:mixedstateprob} we subtract the contribution when a single source is turned off, this implies we have the following,
\begin{align}
\frac{v^{2}}{T^{n_{1}}T^{n_{2}}}\int_{-T/2}^{T/2}\prod_{i_{1}=1}^{n_{1}}dt_{i_{1}}\prod_{i_{2}=1}^{n_{2}}dt_{i_{2}}\Big[\langle n_{d_{1}}(t_{1})n_{d_{2}}(t_{2})\rangle_{n_{1}n_{2}}&-\langle n_{d_{1}}(t_{1})n_{d_{2}}(t_{2})\rangle_{n_{1}}-\langle n_{d_{1}}(t_{1})n_{d_{2}}(t_{2})\rangle_{n_{2}}\Big]\\
&=\frac{n_{1}n_{2}}{T^{2}}P(11)_{\text{CW}}\mp\frac{2n_{1}n_{2}}{T^{2}}\mathcal{F}(x_{1},t_{1};x_{2},t_{2}) \,.
\end{align}
Averaging over the Poisson distributed particles ($n_{1}$ and $n_{2}$), and writing everything in terms of currents, we find,
\begin{equation}
\frac{\langle I_{d_{1}}(t_{1})I_{d_{2}}(t_{2})\rangle_{s_{1}s_{2}}}{\langle I_{s_{1}}\rangle\cdot\langle I_{s_{2}}\rangle}-\frac{\langle I_{d_{1}}(t_{1})I_{d_{2}}(t_{2})\rangle_{s_{1}}}{\langle I_{s_{1}}\rangle\cdot\langle I_{s_{1}}\rangle}-\frac{\langle I_{d_{1}}(t_{1})I_{d_{2}}(t_{2})\rangle_{s_{2}}}{\langle I_{s_{2}}\rangle\cdot\langle I_{s_{2}}\rangle}=P(11)_{\text{CW}}-2\mathcal{F}(x_{1},t_{1};x_{2},t_{2})\,,
\end{equation}
where we have defined the following,
\begin{equation}
\langle I_{s_{1}}\rangle=e\lambda=\langle I_{s_{2}}\rangle\quad(\text{average constant current at the sources \ensuremath{S_{1}} and \ensuremath{S_{2}}})\,.
\end{equation}
Since our detectors are equidistant from the collider but on the opposite sides, Fig.~\ref{fig:extended_collider}a, we have have, $x_{1}+x_{2}=0$. Now consider the equal time cross-current correlator, we obtain the Eq.~\eqref{eq:cross_correlation} of the main text,  
\begin{align}
    \frac{\langle I_{d_{1}}(t)I_{d_{2}}(t)\rangle_{s_{1}s_{2}}}{\langle I_{s_{1}}\rangle\cdot\langle I_{s_{2}}\rangle}-\frac{\langle I_{d_{1}}(t)I_{d_{2}}(t)\rangle_{s_{1}}}{\langle I_{s_{1}}\rangle\cdot\langle I_{s_{1}}\rangle}-\frac{\langle I_{d_{1}}(t)I_{d_{2}}(t)\rangle_{s_{2}}}{\langle I_{s_{2}}\rangle\cdot\langle I_{s_{2}}\rangle}=P(11)_{\text{CW}}-2\mathcal{F}(0) = P(11)_{\text{F}}\,,
\end{align}
where the antibunching probability, $P(11)_{\text{F}}$ is as shown in Fig.~\ref{fig:p11andg2}a,b and also obtained previously in Sec.~\ref{sec:pointandextended}. For extracting the true mutual statistics we would like to write down the benchmark
$\mathcal{B}_{2}$, Eq.~\eqref{eq:irbenchmark} in terms of currents and using Eq.\eqref{eq:singlesourceprob1}-\eqref{eq:singlesourceprob2} we get,
\begin{align}
\mathcal{B}_{2}=\frac{\langle I_{d_{1}}\rangle_{s_{1}}}{\langle I_{s_{1}}\rangle}\frac{\langle I_{d_{2}}\rangle_{s_{2}}}{\langle I_{s_{2}}\rangle}+\frac{\langle I_{d_{1}}\rangle_{s_{2}}}{\langle I_{s_{2}}\rangle}\frac{\langle I_{d_{2}}\rangle_{s_{1}}}{\langle I_{s_{1}}\rangle}=P(11)_{\text{CW}} \,.
\end{align}
Now consider the following with the benchmark,
\begin{align}
P(11)_{\text{F}}-\mathcal{B}_{2} & =P(11)_{\text{F}}-P(11)_{\text{CW}}\\
 & =\frac{\langle I_{d_{1}}(t_{1})I_{d_{2}}(t_{2})\rangle_{s_{1}s_{2}}}{\langle I_{s_{1}}\rangle\cdot\langle I_{s_{2}}\rangle}-\frac{\langle I_{d_{1}}(t_{1})I_{d_{2}}(t_{2})\rangle_{s_{1}}}{\langle I_{s_{1}}\rangle\cdot\langle I_{s_{1}}\rangle}-\frac{\langle I_{d_{1}}(t_{1})I_{d_{2}}(t_{2})\rangle_{s_{2}}}{\langle I_{s_{2}}\rangle\cdot\langle I_{s_{2}}\rangle}\,,\label{eq:crosscurrent1sm}\\
 & \quad-\frac{\langle I_{d_{1}}\rangle_{s_{1}}}{\langle I_{s_{1}}\rangle}\frac{\langle I_{d_{2}}\rangle_{s_{2}}}{\langle I_{s_{2}}\rangle}-\frac{\langle I_{d_{1}}\rangle_{s_{2}}}{\langle I_{s_{2}}\rangle}\frac{\langle I_{d_{2}}\rangle_{s_{1}}}{\langle I_{s_{1}}\rangle}\,,\label{eq:crosscurrent2sm}\\
 & =-2\mathcal{F}(x_{1},t_{1};x_{2},t_{2})\,.
\end{align}
The above is simply a function of $t=t_{1}-t_{2}$ and zero-frequency
Fourier transform of this leads to the following,
\begin{align}
\int_{-T/2}^{T/2}&dt\Bigg[\frac{\langle I_{d_{1}}(t_{1})I_{d_{2}}(t_{2})\rangle_{s_{1}s_{2}}}{\langle I_{s_{1}}\rangle\langle I_{s_{2}}\rangle}-\frac{\langle I_{d_{1}}(t_{1})I_{d_{2}}(t_{2})\rangle_{s_{1}}}{\langle I_{s_{1}}\rangle\langle I_{s_{1}}\rangle}-\frac{\langle I_{d_{1}}(t_{1})I_{d_{2}}(t_{2})\rangle_{s_{2}}}{\langle I_{s_{2}}\rangle\langle I_{s_{2}}\rangle}
-\frac{\langle I_{d_{1}}\rangle_{s_{1}}}{\langle I_{s_{1}}\rangle}\frac{\langle I_{d_{2}}\rangle_{s_{2}}}{\langle I_{s_{2}}\rangle}-\frac{\langle I_{d_{1}}\rangle_{s_{2}}}{\langle I_{s_{2}}\rangle}\frac{\langle I_{d_{2}}\rangle_{s_{1}}}{\langle I_{s_{1}}\rangle}\Bigg] \\ &=-2\int_{-T/2}^{T/2}dt\mathcal{F}(x_{1},t_{1};x_{2},t_{2})\,.
\end{align}
This can be evaluated (assuming $T$ is very large) as follows,
\begin{align}
-2\int_{-T/2}^{T/2}dt\mathcal{F}(x_{1},t_{1};x_{2},t_{2}) & =2\int_{-T/2}^{T/2}dt\Big[\int_{-\infty}^{\infty} dk_{1}e^{ik_{1}(x_{1}+x_{2}+vt_{2}-vt_{1})}\mathcal{T}_{k_{1}}\mathcal{R}_{k_{1}}^{*}|\phi(k_{1})|^{2}\\
 & \quad\times\int_{-\infty}^{\infty} dk_{2}e^{-ik_{2}(x_{1}+x_{2}+vt_{2}-vt_{1})}\mathcal{T}_{k_{2}}^{*}\mathcal{R}_{k_{2}}|\phi(k_{2})|^{2}\Big]\,,\\
 & =2\Big[\int_{-\infty}^{\infty} dk_{1}e^{ik_{1}(x_{1}+x_{2})}\mathcal{T}_{k_{1}}\mathcal{R}_{k_{1}}^{*}|\phi(k_{1})|^{2}\\
 & \quad\times\int_{-\infty}^{\infty} dk_{2}e^{-ik_{2}(x_{1}+x_{2})}\mathcal{T}_{k_{2}}^{*}\mathcal{R}_{k_{2}}\int_{-T/2}^{T/2}dte^{iv(t_{2}-t_{1})[k_{1}-k_{2}]}\,,\\
 & =\frac{4\pi}{v}\Big[\int_{-\infty}^{\infty} dk_{2}dk_{1}e^{ik_{1}(x_{1}+x_{2})}\mathcal{T}_{k_{1}}\mathcal{R}_{k_{1}}^{*}|\phi(k_{1})|^{2}\\
 & \quad\times e^{-ik_{2}(x_{1}+x_{2})}\mathcal{T}_{k_{2}}^{*}\mathcal{R}_{k_{2}}|\phi(k_{2})|^{2}\delta(k_{1}-k_{2})\Big]\,,\\
 & =\frac{4\pi}{v}\int_{-\infty}^{\infty} dk|\mathcal{T}_{k}|^{2}|\mathcal{R}_{k}|^{2}|\phi(k)|^{4}\,.
\end{align}
This implies that the zero-frequency Fourier transform is positive
definite of fermions and a similar analysis gives a negative definite
value of bosons. Extracting quantum statistics by defining such irreducible cross-current correlators Eq.~\eqref{eq:crosscurrent1sm}-\eqref{eq:crosscurrent2sm} have also been used in Ref.~\cite{PhysRevLett.91.196803,zhang2023measuring,zhang_fractional-statistics-induced_2025}.

\section{$g^{(2)}(t_{1},t_{2})$-function and its relation to time-resolved current auto-correlator}\label{sec:g2function}

In this section we are going to calculate the density-density correlations at a given detector, say, $D_{1}$ measured at different times $t_{1}$ and $t_{2}$. In the literature, this is known as the $g^{(2)}-$function~\cite{loudon_quantum_2000}. In the subsection~\ref{subsec:g2particles}, we perform a pedagogical exercise, where we have just two incoming particles from each of the sources, $S_{1}$ and $S_{2}$ and denote the density-density correlator in this case with $g^{(2)}$. In the following subsection~\ref{subsec:g2beam}, we generalize the formalism for a stochastic beam of particles from the sources, $S_{1}$ and $S_{2}$. We perform an averaging for a dilute stochastic beam of particles over the emission times and denote the density-density correlator in this case with $\bar{g}^{(2)}$. We demonstrate how, one can extract the mutual statistics of the colliding particles by studying the zero-frequency Fourier transform of the irreducible part of $\bar{g}^{(2)}$. Also, we demonstrate how the time resolved  $\bar{g}^{(2)}-$function can be used to detect the non-point-like geometry of the colliders. 
\subsection{$g^{(2)}(t_{1},t_{2})-$function for two incoming particles (a pedagogical introduction)}\label{subsec:g2particles}
\begin{figure}[h!]
    \centering
    \includegraphics[width=0.8\linewidth]{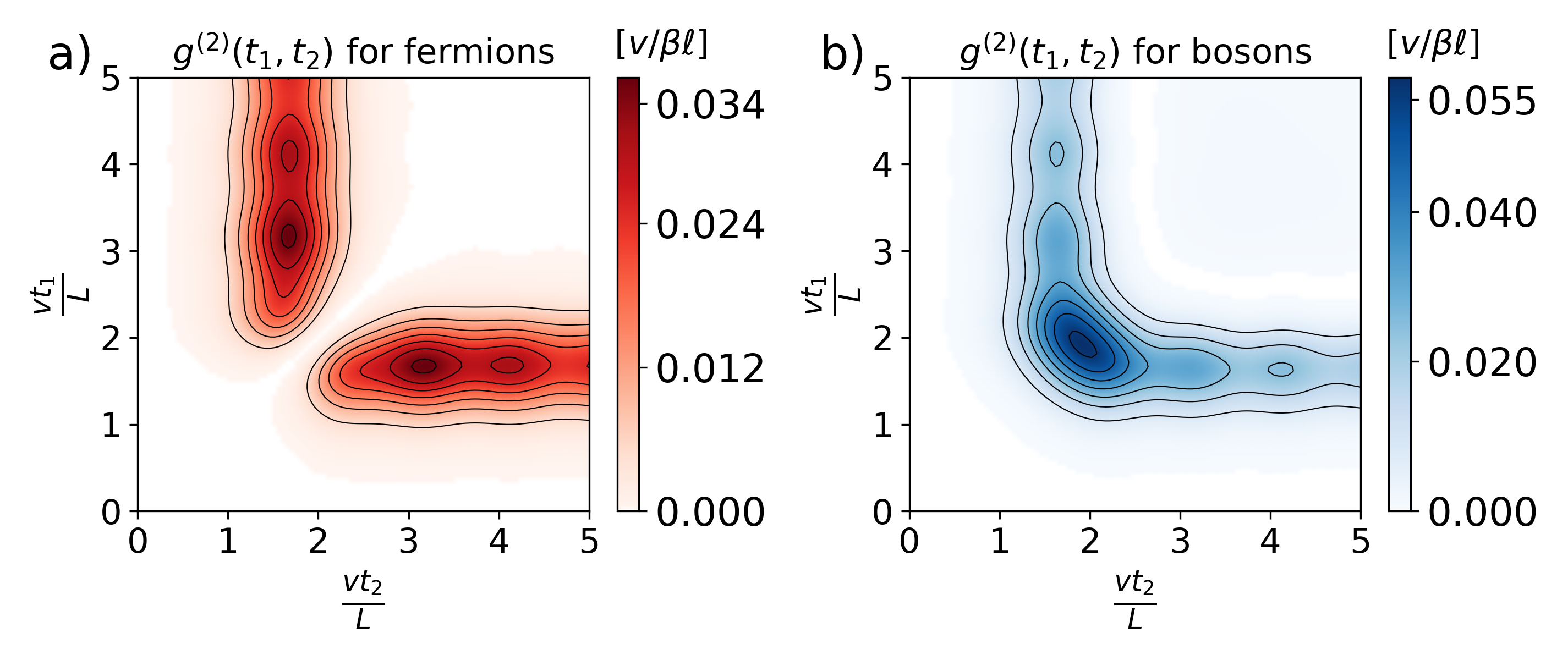}
    \caption{The pair-correlation function $g^{(2)}(t_{1},t_{2})$  [see Eq.~(\ref{eq:g2})]   for fermions/bosons that arrive simultaneously at the collider but are detected at times $t_{1},t_{2}$ in the same drain. The wave packets are emitted simultaneously at $t=0$ and the collider is at a distance $5L/6$ from the sources and the detectors.  
    We use $\frac{\ell}{L}=2.5$, $k_f = 0$, $r=0.95$, $\theta =0$ and a finite temperature $\beta^{-1} = eV/75$, see Sec.~\ref{sec:finitetempwp}. (a)~\textit{Fermions}. The function  $g^{(2)}$ is generally 
    non-vanishing in an extended collider (hence a characteristic feature), unlike in the case of a point-like collider. Vanishing of $g^{(2)}$ at $t_{1}=t_{2}$ validates exclusion property of fermions. (b)~\textit{Bosons}. Non-vanishing along $t_{1}=t_{2}$ shows that bosons can be found in the same location at the same time.  }
    \label{fig:g2particles}
\end{figure}

\begin{figure}[h!]
    \centering
    \subfloat{\includegraphics[width=0.8\textwidth]{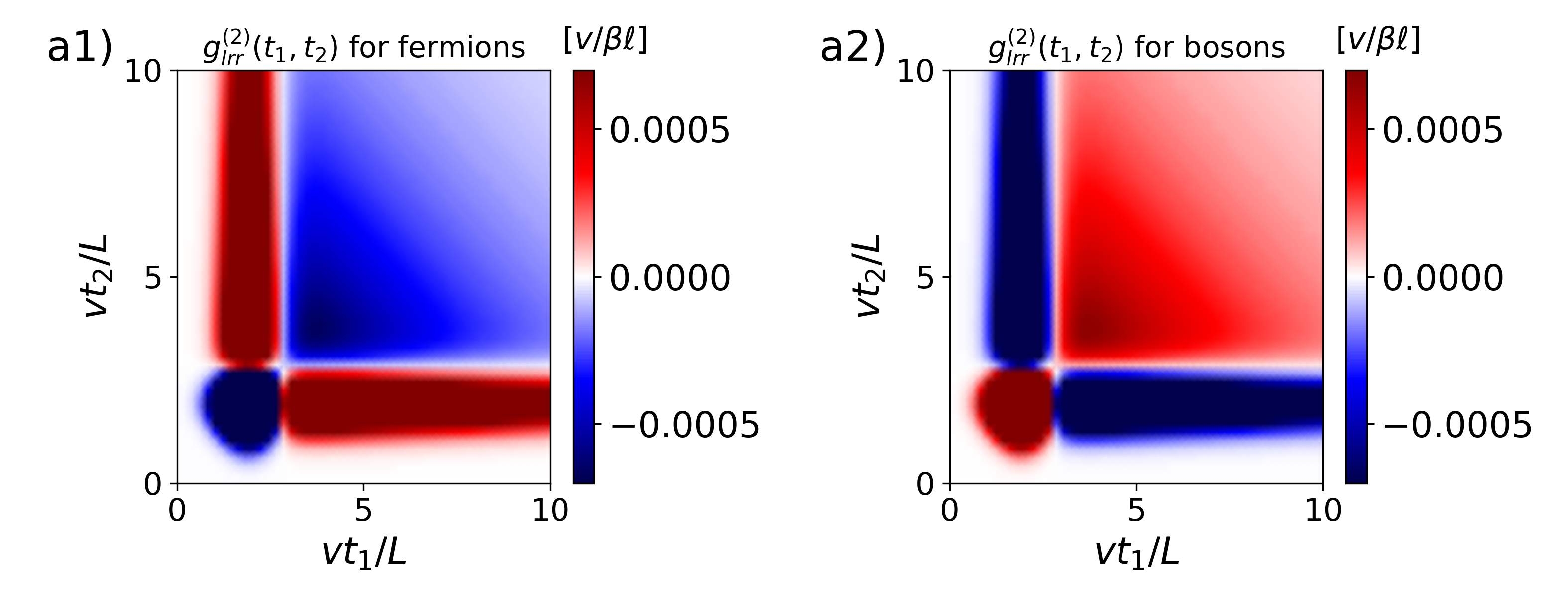}}\\
    \subfloat{\includegraphics[width=0.8\textwidth]{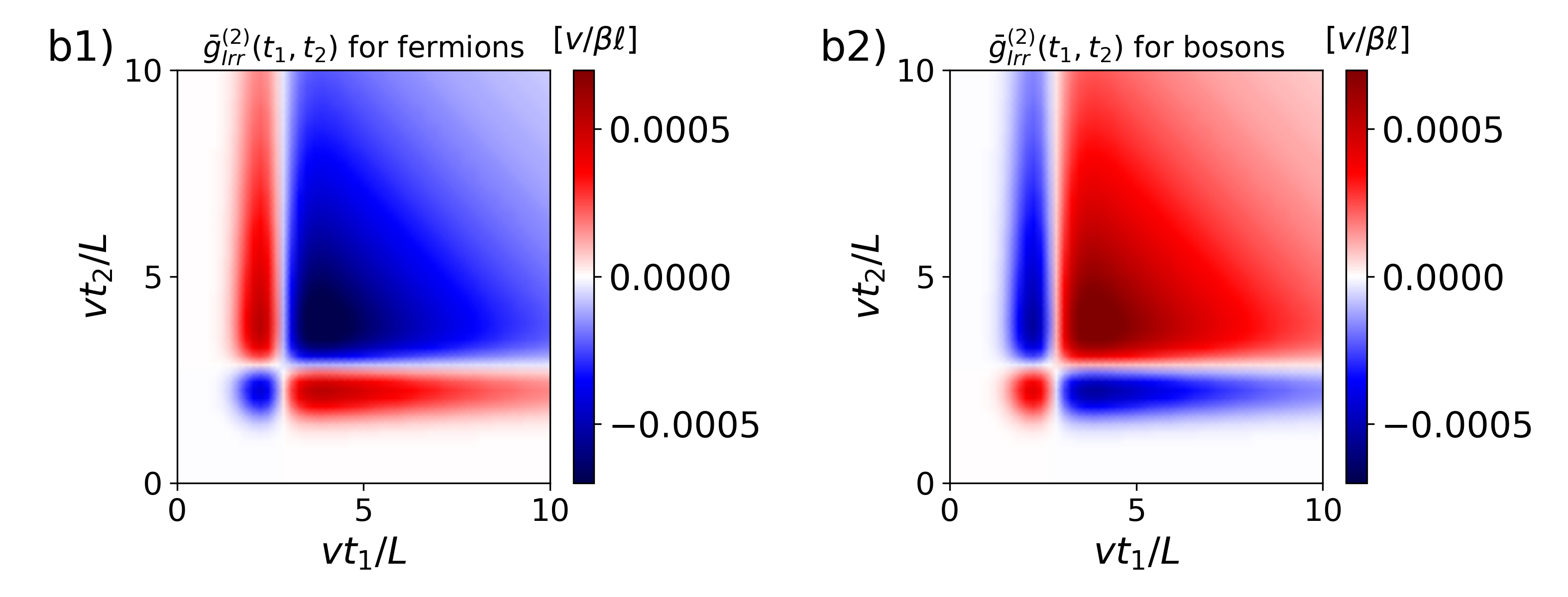}}
    \caption{Plot for $g^{(2)}_{\text{Irr}}-$function for a pair of incoming particles from the sources $S_{1}$ and $S_{2}$. (a)~$g^{(2)}_{\text{Irr}}-$function for a pair of (a1) fermions and (a2) bosons without a delay in the arrival times.  
    (b)~$g^{(2)}_{\text{Irr}}-$function for a pair of (b1) fermions and (b2) bosons with a delay $ L/v$ in the arrival times.  
    }
    \label{fig:g2_reduced}
\end{figure}

Consider, an incoming state with two particles, one from each source. The $g^{(2)}-$function is defined as follows,
\begin{align} \label{eq:g2twoparticle}
    g^{(2)}(t_{1},t_{2}) = v^{2}\langle \psi| n_{d_{1}}(t_{1})n_{d_{1}}(t_{2}) |\psi \rangle\,,
\end{align}
where the expectation value is calculated with respect to the following incoming state,
\begin{align}
    |\psi\rangle = \tilde{\phi}_{s_{1}}^{\dagger}(x_{1}^{(0)},t_{1}^{(0)})\tilde{\phi}_{s_{2}}^{\dagger}(x_{2}^{(0)},t_{2}^{(0)})|\Omega\rangle \,.
\end{align}
From here one can write down the $g^{(2)}(t_{1},t_{2})$ as follows~\cite{Blanter_2000},
\begin{align}
    g^{(2)}(t_{1},t_{2}) = v^{2}|\langle \Omega | \phi(x_{1},t_{1}) \phi(x_{1},t_{2}) \tilde{\phi}_{s_{1}}^{\dagger}(x_{1}^{(0)},t_{1}^{(0)})\tilde{\phi}_{s_{2}}^{\dagger}(x_{2}^{(0)},t_{2}^{(0)}) |\Omega\rangle|^{2}\,.
\end{align}
This implies,
\begin{align}
    g^{(2)}(t_{1},t_{2}) = v^{2}|\mathcal{I}_{1}(x_{1},t_{1};x_{1}^{(0)},t_{1}^{(0)})\mathcal{J}_{2}(x_{1},t_{2};x_{2}^{(0)},t_{2}^{(0)}) \pm \mathcal{I}_{1}(x_{1},t_{2};x_{1}^{(0)},t_{1}^{(0)})\mathcal{J}_{2}(x_{1},t_{1};x_{2}^{(0)},t_{2}^{(0)})|^{2} \,, \label{eq:g2}
\end{align}
where the signs $+/-$ correspond to bosons/fermions 
and the expressions for $\mathcal{I}_{1}$ and $\mathcal{J}_{2}$ appear in the two-particle amplitudes in Eqs.~\eqref{eq:directamp11} and \eqref{eq:exchamp11}. 
The function $g^{(2)}(t_{1},t_{2}) $, Eq.~(\ref{eq:g2}), is plotted in Fig.~\ref{fig:g2particles}c-d under the assumption that 
 the tunneling points to the QAD from the chiral edge states, Fig.~1a,  
are at equal distances from the sources and the emission happens at the same time, $t_{1}^{(0)}=t_{2}^{(0)}$.
Now, let us try to understand the behavior of the $g^{(2)}(t_{1},t_{2})$ in the case where the emission time of the particles are different i.e. $t_{1}^{(0)}\neq t_{2}^{(0)}$. Due to the finite width of the wave packets, interference between the wave packets depends on the emission times of the wave packets. If the wave packets are emitted at very different times, in that case there will no interference in the system and the result will be statistics independent. Consider the following expression for the $g^{(2)}-$function obtained by expanding Eq.~\eqref{eq:g2},
\begin{align}
    g^{(2)}(t_{1},t_{2}) &= v^{2}|\mathcal{I}_{1}(x_{1},t_{1};x_{1}^{(0)},t_{1}^{(0)})|^{2}|\mathcal{J}_{2}(x_{1},t_{2};x_{2}^{(0)},t_{2}^{(0)})|^{2} + v^{2}|\mathcal{I}_{1}(x_{1},t_{2};x_{1}^{(0)},t_{1}^{(0)})|^{2}|\mathcal{J}_{2}(x_{1},t_{1};x_{2}^{(0)},t_{2}^{(0)})|^{2} \,, \\
    &\quad \pm 2v^{2}\text{Re}\big[\mathcal{I}_{1}^{*}(x_{1},t_{1};x_{1}^{(0)},t_{1}^{(0)})\mathcal{J}_{2}^{*}(x_{1},t_{2};x_{2}^{(0)},t_{2}^{(0)})\mathcal{I}_{1}(x_{1},t_{2};x_{1}^{(0)},t_{1}^{(0)})\mathcal{J}_{2}(x_{1},t_{1};x_{2}^{(0)},t_{2}^{(0)})\big] \,, \\
    &\approx v^{2}|\mathcal{I}_{1}(x_{1},t_{1};x_{1}^{(0)},t_{1}^{(0)})|^{2}|\mathcal{J}_{2}(x_{1},t_{2};x_{2}^{(0)},t_{2}^{(0)})|^{2} + v^{2}|\mathcal{I}_{1}(x_{1},t_{2};x_{1}^{(0)},t_{1}^{(0)})|^{2}|\mathcal{J}_{2}(x_{1},t_{1};x_{2}^{(0)},t_{2}^{(0)})|^{2} \,,
\end{align}
where we have used the fact that in the case when the emission time of the wave packets are very different then $\mathcal{I}_{1}^{*}(x_{1},t_{1};x_{1}^{(0)},t_{1}^{(0)})\mathcal{J}_{2}(x_{1},t_{1};x_{2}^{(0)},t_{2}^{(0)})\to 0$, because of the finite width of the incoming wave packets. For fermions, in a point-like collider  we find that $g^{(2)}(t_{1},t_{2})=0$ for all $t_1,t_2$, meaning that the two fermions always end up in different drains, $P(11)_{\text{F}} = 1$. In the extended collider generically $g^{(2)}(t_{1},t_{2})\neq0$ except for simultaneous detection $t_1 = t_2$, since the two fermions cannot occupy the same position simultaneously in the same detector. By integrating Eq.~\eqref{eq:g2}, we obtain the bunching probability,   
\begin{align}
    \frac{1}{2}\iint_{-\infty}^{\infty}dt_{1}dt_{2} g^{(2)}(t_{1},t_{2}) = P(20)_{\text{F}}\,,
\end{align}
which is in general non-zero in an extended collider. In order to extract the mutual statistics from the $g^{(2)}-$function, one can look at the following,
\begin{align}
     g^{(2)}_{\text{Irr}}(t_{1},t_{2}) = v^{2}\langle n_{d_{1}}(t_{1})n_{d_{1}}(t_{2}) \rangle_{s_{1}s_{2}} - v^{2}\langle n_{d_{1}}(t_{1}) \rangle_{s_{1}} \langle n_{d_{1}}(t_{2}) \rangle_{s_{2}} -v^{2}\langle n_{d_{1}}(t_{2}) \rangle_{s_{1}} \langle n_{d_{1}}(t_{1}) \rangle_{s_{2}}\,,
\end{align}
where the subscript refers to the active source(s). This effectively gets rid-off all the single particle self-interference terms and we obtain only statistics dependent part,
\begin{align}\label{eq:greduced}
    g^{(2)}_{\text{Irr}}(t_{1},t_{2}) = \pm 2v^{2}\text{Re}\big[\mathcal{I}_{1}^{*}(x_{1},t_{1};x_{1}^{(0)},t_{1}^{(0)})\mathcal{J}_{2}^{*}(x_{1},t_{2};x_{2}^{(0)},t_{2}^{(0)})\mathcal{I}_{1}(x_{1},t_{2};x_{1}^{(0)},t_{1}^{(0)})\mathcal{J}_{2}(x_{1},t_{1};x_{2}^{(0)},t_{2}^{(0)})\big] \,,
\end{align}
where the +/- signs are for bosons and fermions. However, if the particles are arriving at the collider at very different times then, there will be less interference between the wave packets and therefore $g^{(2)}_{\text{Irr}}(t_{1},t_{2})\to 0$. This result can intuitively be understood from the fact that less interference implies less likely for the particles to wind around each other (exchange processes) and less effect of mutual statistics on the $g^{(2)}_{\text{Irr}}-$function ($|g^{(2)}_{\text{Irr}}|\to 01$), Fig.~\ref{fig:g2_reduced}.  The above presentation assumes that we have two incoming particles from the two different sources, $S_{1}$ and $S_{2}$. 

\subsection{$\bar{g}^{(2)}(t_{1},t_{2})-$function for a stochastic beam of particles}\label{subsec:g2beam}
\begin{figure}[h!]
    \centering
    \includegraphics[width=0.8\linewidth]{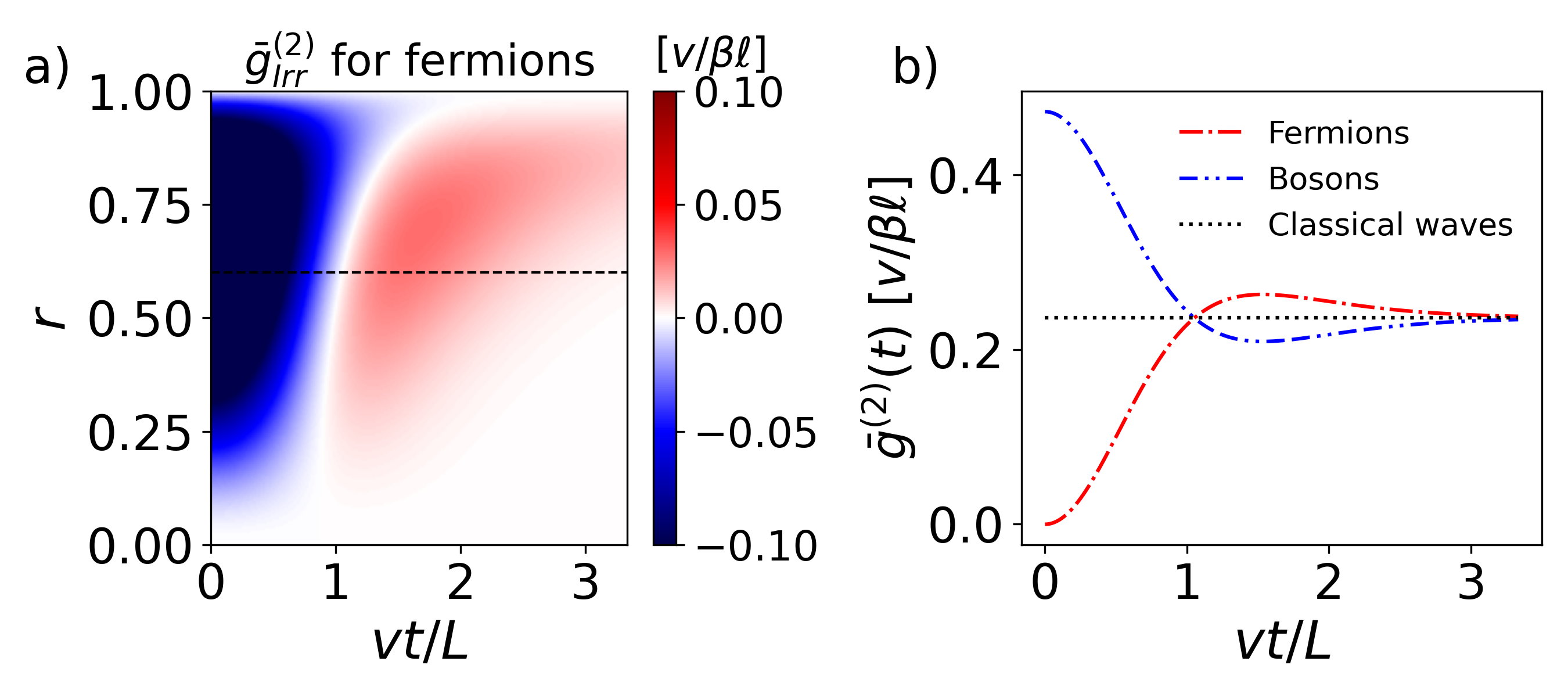}
    \caption{The irreducible part $\bar{g}^{(2)}_{\text{Irr}}(t)$ of the pair-correlation function $\bar{g}^{(2)}(t)$ for a stochastic beam of particles with emission rate $\lambda$ as function of the detection time difference, $t$ (see Eq.~(\ref{eq:g2irrbeam}). 
    (c)~The function $\bar{g}^{(2)}_{\text{Irr}}$ for a dilute stochastic beam of fermions. 
    Due to the extended nature of the collider,  $\bar{g}^{(2)}_{\text{Irr}}(t)$ may change sign as a function of $t$, whereas it has a fixed sign  in a point-like collider. 
    (d)~For bosons, $\bar{g}^{(2)}_{\text{Irr}}$ has the opposite sign from fermions while for classical waves it vanishes.     
    Thus $\bar{g}^{(2)}_{\text{Irr}}$ is sensitive to the mutual statistics and hence can probe the statistics. 
    We use $\frac{\ell}{L}=2.5$, $k_f = 0$, $r=0.6$, $\theta =0$ and a finite temperature $\beta^{-1} = eV/75$, see Sec.~\ref{sec:finitetempwp}. }
    \label{fig:g2avgbeams}
\end{figure}

In this subsection we are going to generalize the above calculation for a stochastic beam of particles. In one-dimensional channel, current at the detectors is proportional to the number density and is given as, $I_{d_{i}}(t_{i})=evn_{d_{i}}(t_{i})$ where $e$ is electronic charge and $v$ is the velocity and $i=1,2$. This implies that the $\bar{g}^{(2)}(t_{1},t_{2})$ function is an auto-correlator of the currents at a given detector averaged over emission times of the stochastic beam of particles from the sources $S_{1}$ and $S_{2}$, Eq.~\eqref{eq:stochasticbeams1s2} (the $\bar{\cdot}$ refers to this averaging),
\begin{align}
    \bar{g}^{(2)}(t_{1},t_{2})=\frac{1}{e^{2}}\langle I_{d_{1}}(t_{1})I_{d_{1}}(t_{2}) \rangle \, .
\end{align} 
[Compare to Eq.~(\ref{eq:g2twoparticle}).] 
Let us assume that the sources are
turned ON for the interval $[-T/2,T/2]$, we assume in the calculations that $T\to\infty$. Again, we start by looking at a particular case where we have $n_{1}$
 particles from $S_{1}$ and $n_{2}$  particles
from source $S_{2}$. By averaging over the incoming times $\{t_{i_{1}}^{(0)},t_{i_{2}}^{(0)}\}$
in the interval $[-T/2,T/2]$, we obtain the following,
\begin{align}
\bar{g}^{(2)}(t_{1},t_{2})_{n_{1}n_{2}} & =\frac{v^{2}}{T^{n_{1}}T^{n_{2}}}\int_{-T/2}^{T/2}\prod_{i_{1}=1}^{n_{1}}dt_{i_{1}}^{(0)}\prod_{i_{2}=1}^{n_{2}}dt_{i_{2}}^{(0)}\Big[\langle n_{d_{1}}(t_{1})n_{d_{1}}(t_{2})\rangle_{n_{1}n_{2}}-\langle n_{d_{1}}(t_{1})n_{d_{1}}(t_{2})\rangle_{n_{1}}-\langle n_{d_{1}}(t_{1})n_{d_{1}}(t_{2})\rangle_{n_{2}}\Big]\,,\\
 & =\frac{2n_{1}n_{2}e^{2}}{T^{2}}P(20)_{\text{CW}}\pm\frac{2n_{1}n_{2}e^{2}}{T^{2}}F(t_{1},t_{2})\label{eq:n1n2particleg2}\,,
\end{align}
where $+/-$ corresponds to bosons and fermions and the subscript
``CW'' refers to the classical waves and the subscript on the LHS tells us that we are working with a particular state, $|\psi_{n_{1}n_{2}}\rangle_{s_{1}s_{2}}$. By subtracting the contribution from the single active source terms, we remove all the possibility where the two particles comes from the same source, a possibility not accounted in the previous sub-sec.~\ref{subsec:g2particles}. The function $F$ is found
to be only depend on the time-difference ($t_{1}-t_{2}$) and is given as follows,
\begin{equation}
F(t_{1}-t_{2})=\text{Re}\Big{[}\int_{-\infty}^{\infty} dk_{1}e^{ik_{1}v(t_{1}-t_{2})}|\phi(k_{1})|^{2}|\mathcal{T}_{k_{1}}|^{2}\int_{-\infty}^{\infty} dk_{2}e^{-ik_{2}v(t_{1}-t_{2})}|\phi(k_{2})|^{2}|\mathcal{R}_{k_{2}}|^{2}\Big{]}\,.
\end{equation}
For a stochastic beam of particles from the sources $S_{1}$ and $S_{2}$
with emission rate $\lambda$, we can take the average over the Poisson distribution of  Eq.~\eqref{eq:n1n2particleg2} and by noting $F(t_{1},t_{2})=F(t_{1}-t_{2})$
we get for a stochastic beam of particles from $S_{1}$ and $S_{2}$,
\begin{equation}\label{eq:g2function}
\bar{g}^{(2)}(t_{1}-t_{2})=2\lambda^{2}P(20)_{\text{CW}}\pm2\lambda^{2}F(t_{1}-t_{2}) = \langle n_{d_{1}}(t_{1})n_{d_{1}}(t_{2}) \rangle_{s_{1}s_{2}}-\langle n_{d_{1}}(t_{1})n_{d_{1}}(t_{2}) \rangle_{s_{1}} - \langle n_{d_{1}}(t_{1})n_{d_{1}}(t_{2}) \rangle_{s_{2}}\,,
\end{equation}
this can further be written as,
\begin{equation}\label{eq:g2functionbeam}
\frac{1}{2\lambda^{2}}\bar{g}^{(2)}(t_{1}-t_{2})=P(20)_{\text{CW}}\pm F(t_{1}-t_{2})\,.
\end{equation}
For extracting the statistics we can define an irreducible $\bar{g}^{(2)}_{\text{Irr}}-$function
by removing the self-interference terms from Eq.~(\ref{eq:g2function}) and this is analogous to the action of subtracting the benchmark, $\mathcal{B}_{2}$ from $P(11)_{\text{F}}$ as done in section, Sec.~\ref{sec:quantumstatistics}. For a particular state $|\psi_{n_{1}n_{2}}\rangle_{s_{1}s_{2}}$, the irreducible $\bar{g}^{(2)}_{\text{Irr}}(t_{1}-t_{2})_{n_{1}n_{2}}$ is given as,
\begin{align}
\bar{g}^{(2)}_{\text{Irr}}(t_{1}-t_{2})_{n_{1}n_{2}} & =\frac{v^{2}}{T^{n_{1}}T^{n_{2}}}\int_{-T/2}^{T/2}\prod_{i_{1}=1}^{n_{1}}dt_{i_{1}}\prod_{i_{2}=1}^{n_{2}}dt_{i_{2}}\Big[\langle n_{d_{1}}(t_{1})n_{d_{1}}(t_{2})\rangle_{n_{1}n_{2}}-\langle n_{d_{1}}(t_{1})n_{d_{1}}(t_{2})\rangle_{n_{1}}-\langle n_{d_{1}}(t_{1})n_{d_{1}}(t_{2})\rangle_{n_{2}}\\
 & \;-\langle n_{d_{1}}(t_{1})\rangle_{n_{1}}\langle n_{d_{1}}(t_{2})\rangle_{n_{1}}-\langle n_{d_{1}}(t_{1})\rangle_{n_{2}}\langle n_{d_{1}}(t_{2})\rangle_{n_{2}}\Big]\,.
\end{align}
In this case, the reducible terms (that includes self interference terms) can be written as follows,
\begin{align}
    \frac{v^{2}}{T^{n_{1}}T^{n_{2}}}\int_{-T/2}^{T/2}\prod_{i_{1}=1}^{n_{1}}dt_{i_{1}}\prod_{i_{2}=1}^{n_{2}}dt_{i_{2}}\big[\langle n_{d_{1}}(t_{1})\rangle_{s_{1}}\langle n_{d_{1}}(t_{2})\rangle_{s_{1}}+\langle n_{d_{1}}(t_{1})\rangle_{s_{2}}\langle n_{d_{1}}(t_{2})\rangle_{s_{2}}\big] = \frac{2n_1 n_2}{T^{2}}\mathcal{P}(1\to1)\mathcal{P}(2\to1)\,,
\end{align}
Taking an average over $n_{1}$ and $n_{2}$, leads to $2\lambda^{2}\mathcal{P}(1\to1)\mathcal{P}(2\to1)=2\lambda^{2}P(20)_{\text{CW}}$. Hence, for a stochastic beam of particles, we are going to have the following
\begin{equation}\label{eq:g2irrbeam}
\frac{1}{2\lambda^{2}}\bar{g}^{(2)}_{\text{Irr}}(t_{1}-t_{2})=\pm F(t_{1}-t_{2})\,.
\end{equation}
where $+/-$ is for bosons and fermions and it is shown in the Fig.~\ref{fig:g2avgbeams}a-b. Zero-frequency Fourier transformation (integration over the time difference
$t=t_{1}-t_{2}$) of the $\bar{g}^{(2)}_{\text{Irr}}-$function leads to
the following,
\begin{equation}
\frac{1}{2\lambda^{2}}\int_{-\infty}^{\infty} dt\bar{g}^{(2)}_{\text{Irr}}(t)=\pm \int_{-\infty}^{\infty}dtF(t)=\pm\frac{2\pi}{v}\int_{-\infty}^{\infty} dk|\mathcal{T}_{k}|^{2}|\mathcal{R}_{k}|^{2}|\phi(k)|^{4}\,,
\end{equation}
which implies that the zero-frequency Fourier transform of irreducible part of the $\bar{g}^{(2)}-$function
is sign definite for bosons (positive) and fermions (negative) and
hence determines the statistics.
Though the zero-frequency Fourier transform of $\bar{g}^{(2)}_{\text{Irr}}$ is sign definite, this is not the case with the $\bar{g}^{(2)}_{\text{Irr}}(t)$ function, Eq.~\eqref{eq:g2irrbeam}. As we vary time, $t$, the sign changes for both bosons and fermions, see Fig.~\ref{fig:g2avgbeams}a-b. This is an intrinsic property of an extended collider and is not the case with the point-like collider. In the case of point-like collider, the transmission/reflection coefficients are independent of momentum i.e. $\mathcal{T}_{k} = \mathcal{T}$ and $\mathcal{R}_{k} = \mathcal{R}$. As a result of this the $\bar{g}^{(2)}_{\text{Irr}}$ which is proportional to $F$ is given as,
\begin{align}
    \frac{1}{2\lambda^{2}}\bar{g}_{\text{Irr}}^{(2)}(t_{1}-t_{2})=\pm F(t_{1}-t_{2})&=\pm \text{Re}\Big{[}\int_{-\infty}^{\infty} dk_{1}e^{ik_{1}v(t_{1}-t_{2})}|\phi(k_{1})|^{2}|\mathcal{T}_{k_{1}}|^{2}\int_{-\infty}^{\infty} dk_{2}e^{-ik_{2}v(t_{1}-t_{2})}|\phi(k_{2})|^{2}|\mathcal{R}_{k_{2}}|^{2}\Big{]}\,, \\    &=\pm \mathcal{T}|^{2}|\mathcal{R}|^{2}\text{Re}\Big{[}\int_{-\infty}^{\infty} dk_{1}e^{ik_{1}v(t_{1}-t_{2})}|\phi(k_{1})|^{2}\int_{-\infty}^{\infty} dk_{2}e^{-ik_{2}v(t_{1}-t_{2})}|\phi(k_{2})|^{2}\Big{]} \,,\\
 &=\pm |\mathcal{T}|^{2}|\mathcal{R}|^{2}\Big{|}\int_{-\infty}^{\infty} dk_{1}e^{ik_{1}v(t_{1}-t_{2})}|\phi(k_{1})|^{2}\Big{|}^{2}\,,
\end{align}
where $+/-$ is for bosons/fermions. Note that the quantity,
\begin{align}
    |\mathcal{T}|^{2}|\mathcal{R}|^{2}\Big{|}\int_{-\infty}^{\infty} dk_{1}e^{ik_{1}v(t_{1}-t_{2})}|\phi(k_{1})|^{2}\Big{|}^{2} \geq 0\,,
\end{align}
and does not change sign as we vary $t=t_{1}-t_{2}$. Hence, the sign change in $\bar{g}^{(2)}_{\text{Irr}}$, Eq.~\eqref{eq:g2irrbeam} for bosons and fermions, Fig.~\ref{fig:g2avgbeams}a-b is a feature that is restricted to an extended collider.

\end{document}